\documentclass[11pt]{article}

\usepackage[margin=1in]{geometry}
\usepackage{amsmath,amssymb,mathtools}
\usepackage{booktabs}
\usepackage{array}
\usepackage{graphicx}
\usepackage{caption}
\usepackage{subcaption}
\usepackage{natbib}
\usepackage{hyperref}
\usepackage[nameinlink,noabbrev]{cleveref}
\usepackage{microtype}
\usepackage{setspace}

\graphicspath{{figures/}}

\hypersetup{hidelinks}

\newcommand{\lat}{\mathrm{lat}}
\newcommand{\lon}{\mathrm{lon}}

\title{\textbf{From Point Anomalies to Anomalous Driving Episodes:}\\
Interpreting Abnormal Driving from Vehicle Telematics}

\author{
Ian Weng Chan\thanks{Nanyang Business School, Nanyang Technological University, Singapore. Email: \texttt{ianweng.chan@ntu.edu.sg}.}
\and
Andrei L. Badescu\thanks{Department of Statistical Sciences, University of Toronto, Canada. Email: \texttt{andrei.badescu@utoronto.ca}.}
}

\date{}

\begin{document}
\maketitle

\begin{abstract}
Vehicle telematics provides detailed observations of driving behaviour that can be useful for automobile insurance risk assessment. Existing approaches often summarize these data through counts of harsh events or individually unusual observations, although driving movements naturally develop over time. Building on a previously proposed hidden Markov model (HMM) framework that identifies pointwise anomalies, we identify short intervals in which a sequence of observations is unusual when considered together. We then examine the original acceleration signals to measure the magnitude and temporal concentration of these episodes and identify recurring patterns of vehicle movement. Temporal concentration describes whether acceleration activity is distributed evenly across an episode or concentrated in relatively few observations.

Using a controlled telematics dataset containing normal, aggressive, and drowsy driving by the same drivers, we evaluate the proposed measures through comparisons matched by driver and road type. We find that statistical anomaly measures depend on the driving population used to train the HMM. More importantly, mean temporal concentration alone distinguishes normal from abnormal driving more effectively than the original pointwise anomaly indices across different reference populations. The recurring movement patterns further reveal different characteristics of abnormal driving, with stronger longitudinal acceleration during aggressive driving and more frequent lateral corrections or excursions during drowsy driving. The framework therefore goes beyond pointwise anomaly detection, providing interpretable behavioural features that can be calculated for unlabeled trips, used for trip-level assessment, aggregated across a driver's trip history, and subsequently evaluated alongside existing telematics and rating variables for insurance risk modelling and ratemaking.
\end{abstract}

\noindent\textbf{Keywords:} vehicle telematics; anomaly detection; hidden Markov model; driving episodes; driving behaviour; usage-based insurance.

\newpage

\section{Introduction}
\label{sec:introduction}

Traditional rating variables and past claims provide useful information about automobile insurance risk, but only indirectly reflect how a policyholder actually drives. Claims are also relatively infrequent, so drivers with similar claim histories may exhibit very different driving behaviour. Vehicle telematics offers a more direct view of how vehicles move on the road through high-frequency records of speed, acceleration, braking, turning, time of day, and mileage. The challenge is then to turn the many observations collected over individual trips into a manageable set of behavioural measures for risk assessment. A growing actuarial literature has shown that telematics information can improve claim-frequency modelling, credibility assessment, and dynamic pricing \citep{Verbelen2018,Denuit2019,Henckaerts2022}, while \citet{Chan2025NAAJ} illustrates how raw telematics signals can be converted into interpretable behavioural summaries before they enter an insurance model.

Many practical summaries are based on isolated events, such as harsh braking, harsh acceleration, hard cornering, or speeding. Related actuarial work has interpreted such events as telematics-based near misses and studied their relation to driving risk and insurance pricing \citep{Guillen2020,Guillen2021,Chan2025NAAJ}. Near-miss events are attractive because they occur much more frequently than insurance claims and can therefore provide more immediate information about driving behaviour. However, a fixed threshold necessarily reduces a short driving sequence to whether one observation crossed a prescribed level. A hard brake may reflect poor anticipation, but it may also be an appropriate response that avoids an accident. Conversely, a driver may repeatedly produce unusual sequences of acceleration or lateral movement without any single observation being sufficiently extreme to trigger a harsh-event rule. For high-frequency telematics, the sequence around an event can therefore be as important as the extreme observation itself.

Event counts and other trip-level summaries reduce the data to a manageable scale, but discard much of the structure in the underlying time series. Driving behaviour evolves over time, and lateral and longitudinal movements often need to be interpreted together. Analysing telematics as a multivariate time series can therefore retain temporal and cross-channel information before constructing trip-level features \citep{Chan2025ASTIN}. Related work has also sought to retain richer local information from high-frequency telematics signals rather than relying only on aggregate trip summaries \citep{Lee2026}.

Our starting point is the hidden Markov model (HMM) anomaly framework of \citet{Chan2025ASTIN}. An HMM is fitted to multivariate telematics trips and describes the evolution of the observations through a collection of latent driving regimes. At each observation time, the fitted model gives a predictive distribution for the next telematics observation conditional on the preceding observations in the trip. Forecast pseudo-residuals then measure how surprising the realized observation is relative to this predictive distribution. This moves anomaly assessment away from a fixed threshold in physical units: an acceleration value is considered unusual because it is unexpected given the fitted driving dynamics and recent history, not simply because it exceeds a predetermined acceleration level. The anomaly index in \citet{Chan2025ASTIN} summarizes observations with sufficiently extreme predictive residuals, analogously to identifying residuals several standard deviations from their reference distribution.

The pointwise index, however, treats each extreme residual separately. Several consecutive exceedances may reflect one driving episode, while a sequence of moderately unusual residuals may be abnormal collectively even if none exceeds the pointwise threshold. The index also counts all exceedances equally, without distinguishing how far they depart from the reference; \citet{Lee2026} similarly highlights the limitations of frequency-based summaries. This leads to the idea of a \emph{collective anomaly}: a departure that becomes apparent when the observations are considered together. This distinction matters for driving, since braking, acceleration, lateral correction, and recovery develop over several observations rather than occurring at an isolated instant.

Methods for collective anomaly detection provide a natural starting point. Collective and Point Anomalies (CAPA) separates isolated anomalies from anomalous intervals and allows the local mean and variance to change \citep{Fisch2022CAPA}; multivariate CAPA (MVCAPA) extends the framework to multivariate sequences and subsets of affected components \citep{Fisch2022MVCAPA}. Maximally Divergent Intervals (MDI) instead searches for intervals whose local distribution differs from a reference distribution \citep{Barz2019}. We adapt this interval-divergence idea to HMM forecast residuals, comparing candidate intervals with a common HMM-generated reference and calibrating the complete interval search by simulation. The calibration accounts for the many locations and interval lengths examined simultaneously; see \citet{WaltherPerry2022} for the broader scan-statistic issue.

Once an anomalous interval has been identified, a different question arises: what kind of driving movement does it represent, and does a similar movement recur? This connects our work to the time-series motif literature. Dynamic time warping (DTW) compares trajectories while allowing modest differences in local timing \citep{SakoeChiba1978}, while motif methods search a time series for recurring subsequences \citep{Lin2007,SchäferLeser2022}. In vehicle telematics, \citet{SilvaHenriques2020} and \citet{SilvaHenriques2021} apply motif discovery to UAH-DriveSet and show that repeated acceleration trajectories can reveal interpretable driving manoeuvres. More recently, \citet{SilvaEtAl2026} study recurring patterns in multivariate residuals. Our use of recurrence is more targeted: rather than searching the complete trip for unknown motifs, we group only the anomalous episodes already localized by the interval detector. The resulting DTW clusters represent recurring episode trajectories and are then assigned to broader physical families based on the acceleration patterns of their medoids.

We therefore propose an episode-based extension of the HMM anomaly framework. The procedure first searches the HMM forecast residuals for short intervals whose local mean and covariance differ from an HMM-generated reference, and retains distinct non-overlapping episodes after simulation-based calibration. We then return to the original lateral and longitudinal acceleration signals to describe each selected episode. In particular, we record its duration and acceleration magnitude, distinguish lateral from longitudinal movement, and introduce temporal concentration (TC) to measure whether the acceleration activity is distributed evenly across an episode or concentrated in relatively few observations. Finally, episodes with similar acceleration trajectories are grouped into recurring DTW clusters and assigned to broader physical families. The output is not one optimized score, but a small collection of complementary quantities describing how often anomalous episodes occur, how strongly they depart from the statistical reference, and how they are physically expressed.

Because claim-linked insurance telematics data are not available for the present study, we evaluate the framework using UAH-DriveSet \citep{Romera2016}, which contains controlled normal, aggressive, and drowsy driving by the same drivers on common road types. This allows us to compare changes in driving condition within the same driver and road setting. Driving-condition labels are used to define alternative HMM reference populations and, after the features have been constructed, to evaluate the measures and fit exploratory classifiers. Importantly, they are not used to detect or characterize episodes, construct DTW clusters, or assign physical families.

The results provide three main findings. First, statistical anomaly burden depends substantially on the HMM reference population. A normal-only reference gives the clearest matched abnormal--normal contrasts, but larger statistical departures do not necessarily lead to better classification. Second, mean TC provides relatively consistent separation between abnormal and normal driving across different references. In an exploratory leave-one-driver-out classification, TC alone gives a higher mean within-driver area under the receiver operating characteristic curve (ROC-AUC) and a lower Brier score than the original pointwise indices, with particularly clear improvements for drowsy driving. Third, the physical families provide further interpretation of the detected abnormality: aggressive driving tends to involve stronger longitudinal episodes, whereas drowsy driving shows more recurring lateral corrections or excursions. These physical profiles help us understand the different forms of abnormal driving behaviour, beyond what can be described by a single anomaly measure.

For actuarial applications, these measures can be calculated for new trips without driving-condition labels and summarized across a driver's repeated driving experience. TC provides a compact measure of abnormal driving, while episode frequency, strength, acceleration magnitude, and physical-family rates retain information about how it is expressed. For ratemaking, however, the present controlled dataset does not contain insurance claim outcomes from which to estimate risk coefficients or pricing relativities. Whether these measures improve risk assessment beyond existing telematics summaries and conventional rating variables therefore needs to be established using claim-linked data.

The remainder of the paper is organized as follows. \Cref{sec:method} develops the HMM residual representation and anomalous-interval detector. \Cref{sec:measures} defines the episode- and trip-level measures and describes the DTW clusters and physical-family assignments. \Cref{sec:data} introduces UAH-DriveSet and the empirical design. \Cref{sec:results} presents the detector validation, matched behavioural comparisons, reference-population sensitivity, exploratory classification, and physical-family profiles. \Cref{sec:conclusion} concludes and discusses the actuarial implications.

\section{Identifying and representing anomalous driving episodes}
\label{sec:method}

The first part of the framework converts the HMM forecast residuals into distinct anomalous driving episodes. The basic idea is to move from asking whether an individual observation is unusually large to asking whether a short sequence of observations is collectively unusual relative to what the fitted HMM would normally generate. Once such an interval has been identified, it also gives us a specific segment of the original acceleration signals that can later be described physically and compared with recurring patterns in other trips.

\subsection{Recap: HMM and forecast pseudo-residuals}
\label{subsec:hmm}

For trip $j$, let
\[
Y_{jt}
=
\begin{pmatrix}
a^{\lat}_{jt}\\
a^{\lon}_{jt}
\end{pmatrix},
\qquad t=1,\ldots,T_j,
\]
denote lateral and longitudinal acceleration. We model each trip as a multivariate time series using an HMM with latent state $S_{jt}\in\{1,\ldots,K\}$. The latent states evolve according to a transition matrix $P=(p_{rs})$, while conditional on state $k$ the two acceleration channels follow state-dependent Normal distributions,
\[
a^d_{jt}\mid S_{jt}=k \sim N(\mu_{kd},\sigma_{kd}^2), \qquad d\in\{\lat,\lon\}.
\]

The HMM is fitted jointly to the training-trip time series. As in \citet{Chan2025ASTIN}, we do not interpret individual hidden states as predefined manoeuvres such as braking, turning, or lane changing. Rather, the states provide a flexible representation of changing acceleration regimes and, most importantly here, a predictive distribution for the next observation.

Let
\[
\omega_{jt}(k)=\Pr(S_{jt}=k\mid\mathcal F_{j,t-1})
\]
be the predicted probability of state $k$ before observation $t$ is seen. The one-step predictive distribution for channel $d$ is
\begin{equation}
F_{jt,d}(y\mid\mathcal F_{j,t-1})
=
\sum_{k=1}^K
\omega_{jt}(k)
\Phi\!\left(
\frac{y-\mu_{kd}}{\sigma_{kd}}
\right),
\label{eq:hmm-predictive}
\end{equation}
where $\mathcal F_{j,t-1}$ contains the preceding observations in the trip. Hence, before observing the acceleration at time $t$, the HMM uses the preceding trip history to assign probabilities to the possible latent states. The predictive distribution in \cref{eq:hmm-predictive} then averages the state-specific acceleration distributions using these probabilities.

We transform the realized observation through its predictive distribution,
\begin{equation}
u_{jt,d}
=
F_{jt,d}\!\left(a^d_{jt}\mid\mathcal F_{j,t-1}\right),
\qquad
z_{jt,d}
=
\Phi^{-1}(u_{jt,d}).
\label{eq:pseudo-residual}
\end{equation}
Under a correctly specified predictive model, each $z_{jt,d}$ is marginally standard Normal. A large positive or negative pseudo-residual therefore means that the observed acceleration is unusual relative to what the HMM predicted from the preceding driving history.

The anomaly index proposed in \citet{Chan2025ASTIN} uses this pointwise idea directly. For acceleration channel $d$, there are $T_j-1$ forecast pseudo-residuals because the first observation in the trip has no preceding observation history. The pointwise anomaly index is therefore
\begin{equation}
A^{\mathrm{point}}_{j,d}
=
\frac{1}{T_j-1}
\sum_{t=2}^{T_j}
\mathbf 1\!\left\{
|z_{jt,d}|\geq 3
\right\}.
\label{eq:point-index}
\end{equation}
Thus $A^{\mathrm{point}}_{j,d}$ is the proportion of observations in trip $j$ that are individually extreme relative to the fitted HMM.

This pointwise index provides a simple and interpretable measure of extreme local departures, but it does not use the surrounding sequence. Several moderately unusual residuals may together form a meaningful departure even when none individually exceeds three standard deviations; conversely, several consecutive extreme residuals may all arise from one continuous driving event. The present paper therefore changes the unit of analysis from individual residuals to short intervals of the bivariate residual sequence
\[
z_{jt}
=
(z_{jt,\lat},z_{jt,\lon})^\top.
\]
The purpose is not only to detect that something unusual has occurred, but also to identify where the unusual segment begins and ends so that we can return to the original acceleration signals and examine what kind of driving movement occurred during that episode.

\subsection{From pseudo-residuals to anomalous episodes}
\label{subsec:intervals}

The interval detector has three main steps. First, we form short candidate intervals from the forecast pseudo-residuals and measure how different each interval is from a common HMM-generated reference. Second, because many overlapping intervals are searched, we determine how large such a score could become by chance over the whole search. Finally, we select a set of distinct, non-overlapping intervals with sufficient evidence of departure from the reference. The interval score is motivated by the maximally divergent intervals (MDI) approach of \citet{Barz2019}, while the final penalized selection of distinct episodes is related to CAPA \citep{Fisch2022CAPA}.

\paragraph{Building candidate intervals from the residuals.}
We first construct a $q$-step \emph{residual embedding} by stacking $q$ consecutive bivariate forecast residuals,
\begin{equation}
z^{(q)}_{jt}
=
\left(
z_{j,t-q+1}^{\top},
\ldots,
z_{jt}^{\top}
\right)^{\top}
\in\mathbb R^{2q}.
\label{eq:embedding}
\end{equation}

The main analysis uses $q=1$, so $z^{(1)}_{jt}=z_{jt}$ contains only the current lateral and longitudinal pseudo-residuals. With $q=2$, for example, each position contains the current residual pair together with the previous pair. Larger $q$ therefore allows the detector to respond to unusual short-lag relationships in addition to changes in the contemporaneous residual distribution. Because the HMM itself already models temporal dependence, we use $q=1$ as the primary specification in the present analysis. The formulas below are therefore written for $q=1$.

A candidate interval $I=[a,b]$ refers to residual positions $a$ through $b$ in the trip and contains $n_I=b-a+1$ bivariate residual vectors. We restrict attention to a prespecified set of short interval lengths. For each allowed length, the interval is moved along the residual sequence one observation at a time, so neighbouring candidates can overlap substantially. The specific durations used in the empirical analysis are given in \cref{sec:data}. We then ask whether the residual vectors within each candidate interval, when considered together, look different from what the fitted HMM would normally generate. In particular, a run of moderately unusual residuals can provide evidence of abnormality even when no individual residual crosses the earlier $\pm3$ threshold.

\paragraph{Constructing the HMM-generated reference.}
Under a well-fitted HMM, each pseudo-residual component is marginally standard Normal. However, we do not simply assume that the bivariate residual process consists of independent $N_2(0,I_2)$ observations. Instead, we construct the reference directly from the fitted HMM. Complete trips are simulated from the model, the simulated latent states are discarded, and the simulated observations are passed through the same filtering and pseudo-residual calculation used for the observed trips. The resulting residual vectors are pooled across the complete simulation bank to form one common reference for that fitted HMM, rather than estimating a separate reference from each observed trip.

This differs from the usual MDI construction in an important way. MDI compares the distribution inside a candidate interval with a background distribution formed from observations outside that interval. For our purpose, the remainder of the same trip is not necessarily an appropriate reference: if a substantial part of a trip is itself unusual, another unusual segment may no longer appear very different from the rest of that trip. We instead ask whether each candidate interval differs from driving dynamics generated by the fitted HMM.

Let $\widehat\mu_0$ and $\widehat\Sigma_0$ denote the mean and covariance of the pooled simulated bivariate residuals, with $\widehat\Sigma_0=L_0L_0^\top$. We standardize
\begin{equation}
X_{jt}
=
L_0^{-1}(z_{jt}-\widehat\mu_0).
\label{eq:whiten}
\end{equation}

After this transformation, the HMM-generated reference has mean zero and covariance $I_2$ in its first two moments. The transformation therefore puts the two residual channels on a common reference scale without assuming that the complete residual sequence consists of independent bivariate Normal observations.

\paragraph{Scoring one candidate interval.}
For a candidate interval $I=[a,b]$, let $\overline X_I$ be the average of its $n_I$ transformed residual vectors and let $S_I$ be their sample covariance matrix. The mean $\overline X_I$ describes whether the residuals in the interval tend to remain systematically away from the HMM reference. The covariance $S_I$ describes their local variability and whether the lateral and longitudinal residuals tend to move together.

Because a short interval contains relatively few observations, its sample covariance can be noisy or nearly singular. More generally, for a $p$-dimensional residual vector and a $q$-step embedding, whose dimension is $pq$, we use the working shrinkage rule
\[
\gamma_I^{(q)}
=
\frac{pq}{n_I+pq}.
\]
Thus shorter intervals and higher-dimensional embeddings receive more shrinkage toward the HMM reference covariance. In our bivariate $q=1$ specification, $pq=2$, giving
\begin{equation}
V_I
=
(1-\gamma_I)S_I+\gamma_I I_2,
\qquad
\gamma_I=\frac{2}{n_I+2}.
\label{eq:shrink}
\end{equation}
The shrinkage decreases as $n_I$ increases, so longer intervals rely increasingly on their own sample covariance. This is a simple working regularization rule rather than an optimal covariance estimator.

We measure how strongly interval $I$ differs from the HMM reference by
\begin{equation}
G(I)
=
n_I
\left[
\underbrace{\|\overline X_I\|^2}_{\text{mean departure}}
+
\underbrace{\operatorname{tr}(V_I)-2-\log\det(V_I)}_{\text{covariance departure}}
\right],
\label{eq:g}
\end{equation}

where $\|\cdot\|$ denotes the Euclidean norm, so
$\|\overline X_I\|^2
=
\overline X_{I,\lat}^2+\overline X_{I,\lon}^2$.
The first component of $G(I)$ becomes large when the residuals remain systematically shifted away from their expected level. The second becomes large when their variability or lateral--longitudinal relationship differs from the reference. Multiplication by $n_I$ means that a departure that persists over a longer interval provides more evidence than the same average departure observed only briefly. Thus, an interval can receive a large score even when every individual residual remains within $\pm3$.

$G(I)$ also has a direct statistical interpretation as a Gaussian Kullback--Leibler (KL) divergence. For two $p$-dimensional Normal distributions,
\begin{equation}
D_{\mathrm{KL}}
\!\left\{
N_p(\mu_1,\Sigma_1)
\,\middle\|\,
N_p(\mu_0,\Sigma_0)
\right\}
=
\frac{1}{2}
\left[
\operatorname{tr}(\Sigma_0^{-1}\Sigma_1)
+
(\mu_0-\mu_1)^\top
\Sigma_0^{-1}
(\mu_0-\mu_1)
-p
+
\log\frac{\det(\Sigma_0)}{\det(\Sigma_1)}
\right].
\label{eq:gaussian-kl}
\end{equation}

Taking $\mu_1=\overline X_I$, $\Sigma_1=V_I$, $\mu_0=0$, $\Sigma_0=I_2$, and $p=2$ gives
\begin{equation}
D_{\mathrm{KL}}
\!\left\{
N_2(\overline X_I,V_I)
\,\middle\|\,
N_2(0,I_2)
\right\}
=
\frac{1}{2}
\left[
\|\overline X_I\|^2
+
\operatorname{tr}(V_I)
-2
-\log\det(V_I)
\right].
\label{eq:g-kl}
\end{equation}

Hence,
\begin{equation}
G(I)
=
2n_I
D_{\mathrm{KL}}
\!\left\{
N_2(\overline X_I,V_I)
\,\middle\|\,
N_2(0,I_2)
\right\}.
\label{eq:g-kl-score}
\end{equation}

The score can therefore be viewed as an MDI-type measure of how different the first two moments within the candidate interval are from the HMM-generated reference. The Gaussian KL expression is used as a convenient discrepancy measure; it does not require the true residual sequence to be jointly Gaussian or imply that every possible form of distributional change will be detected. With $q=1$, the score responds primarily to changes in residual level, variability, and contemporaneous lateral--longitudinal dependence.

\paragraph{Accounting for the many intervals searched.}
A large value of $G(I)$ is not sufficient by itself to declare an anomalous episode. Because intervals of several lengths are moved across the trip, many overlapping candidates are examined. Even for a trip generated exactly from the fitted HMM, the largest score among all these candidates can be fairly large simply because we have searched in many places. The threshold therefore needs to account for the \emph{whole search}, rather than treating each candidate interval separately.

For an observed trip $j$ being scored, let $\mathcal I_j$ denote all candidate intervals produced by its usable residual length and the prespecified interval lengths. We simulate independent calibration trips from the fitted HMM with the same search design and pass each one through exactly the same residual transformation and interval scan. In calibration replicate $b$, we record only the largest score

\begin{equation}
M_j^{(b)}
=
\max_{I\in\mathcal I_j}
G^{(b)}(I),
\qquad b=1,\ldots,B.
\label{eq:calibration-max}
\end{equation}

The maximum is the relevant quantity because we want to know whether any interval in an HMM-generated trip could look this unusual after the complete search. Recording one maximum from each simulated trip does not restrict the observed trip to having only one detected episode.

We set the search penalty $\lambda_j$ to the empirical 95th percentile of $M_j^{(1)},\ldots,M_j^{(B)}$, using 95\% as a prespecified whole-search false-report level. Thus, under the fitted HMM reference, only about 5\% of otherwise ordinary simulated trips produce at least one candidate interval above this level. A larger search generally leads to a larger penalty $\lambda_j$ because there are more opportunities for a large score to occur by chance. The subscript $j$ therefore reflects the search design of trip $j$, rather than a penalty estimated from the observed residual values of that trip.

In the empirical analysis, we use $B=1{,}999$ calibration replicates. Together with the observed scan maximum, this gives 2,000 possible Monte Carlo ranks and hence a whole-search $p$-value resolution of $1/2000$. Two observed trips fitted under the same HMM and having the same usable residual length and candidate-interval design can use the same calibrated value of $\lambda_j$.

\paragraph{Selecting non-overlapping episodes.}
After calibration, many high-scoring candidate intervals may overlap because slightly different intervals can describe the same underlying departure. We therefore select a non-overlapping collection rather than report every high-scoring candidate. Specifically, among all collections $\mathcal E$ of mutually non-overlapping candidate intervals, we choose the collection that maximizes
\begin{equation}
\sum_{I\in\mathcal E}
\{G(I)-\lambda_j\}.
\label{eq:episode-selection}
\end{equation}
We denote the selected collection by $\widehat{\mathcal E}_j$. The empty collection is allowed, so a trip may contain no detected episode, and the number of episodes is not fixed in advance. An interval with $G(I)\leq\lambda_j$ cannot improve the objective and is therefore not selected. We solve \cref{eq:episode-selection} exactly by dynamic programming (weighted interval scheduling). For each candidate interval, we compare the best solution excluding it with its net score plus the best solution ending before it starts. Backtracking then recovers the optimal non-overlapping collection without enumerating all possible combinations.

The penalty also discourages one broad departure from being split unnecessarily into several shorter intervals: every additional interval must provide enough extra evidence to overcome another $\lambda_j$. The non-overlap requirement prevents overlapping windows from repeatedly counting the same departure, while two sufficiently separated departures can both be retained. This construction is related to CAPA \citep{Fisch2022CAPA}, while the local evidence for each candidate interval comes from the HMM-reference divergence in \cref{eq:g}. We refer to the intervals in $\widehat{\mathcal E}_j$ as \emph{anomalous episodes}.

\begin{figure}[h!]
\centering
\includegraphics[width=\textwidth]{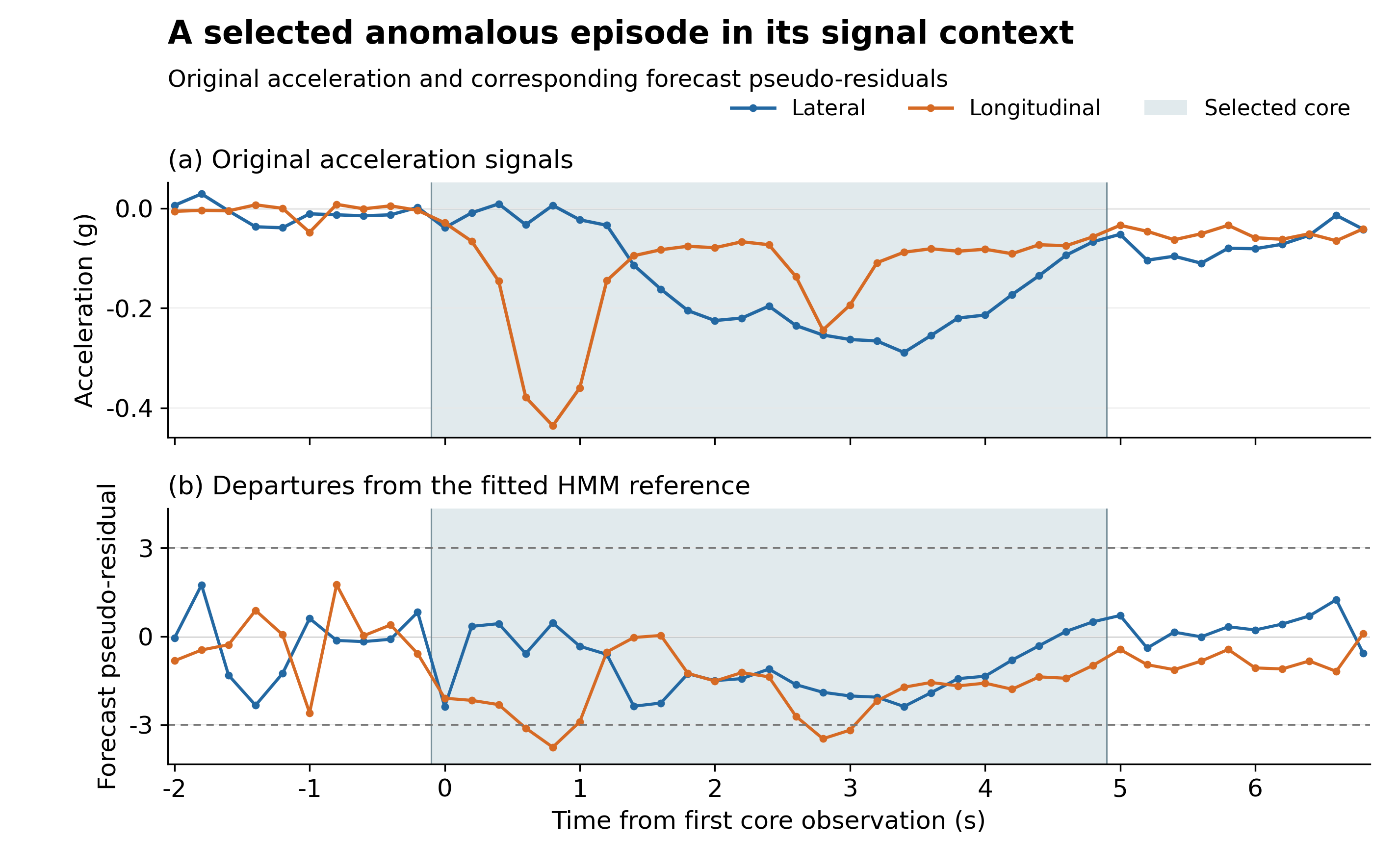}
\caption{Illustration of a selected anomalous driving episode. The upper panel shows the original lateral and longitudinal acceleration, while the lower panel shows the corresponding HMM forecast pseudo-residuals. The dashed horizontal lines indicate the pointwise anomaly thresholds at $\pm3$. The shaded region is the selected five-second anomalous interval. Observations outside the shaded region provide context and are not included in the interval score $G(I)$.}
\label{fig:episode-context}
\end{figure}

The difference from the earlier pointwise anomaly index is illustrated in \cref{fig:episode-context}. Whereas the pointwise index counts individually extreme pseudo-residuals, the interval procedure considers the residuals collectively and identifies a contiguous anomalous interval. In this example, only four of the 25 observations within the selected interval exceed the pointwise threshold of three in either acceleration channel. Once this interval has been selected, we return to the corresponding lateral and longitudinal acceleration to examine the physical driving pattern. The selected interval identifies the statistically anomalous portion of the sequence and need not coincide exactly with a complete physical manoeuvre.

\section{Characterizing and summarizing anomalous driving episodes}
\label{sec:measures}

The procedure in \cref{sec:method} identifies intervals in which the forecast residuals depart collectively from the fitted HMM reference. Once these anomalous episodes have been located, we ask a different question: how should they be summarized, and what do they look like in the original acceleration signals? We first separate how often anomalous episodes occur from how strongly they depart from the statistical reference. We then return to the lateral and longitudinal acceleration trajectories to describe the acceleration magnitude and temporal structure of individual episodes, before grouping recurring trajectories into DTW clusters and assigning them to physical families.

Because the detected anomalous episodes depend on the reference population used to fit the HMM, the resulting trip-level summaries also depend on that reference. We write the reference $R$ explicitly for the statistical quantities below and suppress it from the signal-based quantities when only one reference is under discussion.

\subsection{Statistical summaries of detected episodes}
\label{subsec:statistical-summaries}

Let $E_j$ denote the valid driving exposure of trip $j$ in hours and
\[
N_j(R)
=
\left|\widehat{\mathcal E}_j(R)\right|
\]
the number of detected episodes under reference population $R$. The most direct occurrence measure is the exposure-adjusted episode frequency
\begin{equation}
F_j(R)
=
\frac{N_j(R)}{E_j}.
\label{eq:episode-frequency}
\end{equation}
Thus $F_j(R)$ measures how often distinct anomalous episodes occur per driving hour.

Frequency alone does not distinguish many relatively mild episodes from a smaller number of episodes that depart much more strongly from the reference. For a trip with $N_j(R)>0$, we therefore define the mean excess statistical evidence per detected episode as
\begin{equation}
\overline G^{\,\mathrm{exc}}_j(R)
=
\frac{1}{N_j(R)}
\sum_{e\in\widehat{\mathcal E}_j(R)}
\{G(e;R)-\lambda_j(R)\}.
\label{eq:episode-strength}
\end{equation}
We refer to $\overline G^{\,\mathrm{exc}}_j(R)$ as the \emph{statistical episode strength}. It measures, on average, how far a selected episode exceeds the whole-search admission threshold $\lambda_j(R)$.

Combining frequency and strength gives the exposure-adjusted statistical anomaly burden
\begin{equation}
B_j(R)
=
\frac{1}{E_j}
\sum_{e\in\widehat{\mathcal E}_j(R)}
\{G(e;R)-\lambda_j(R)\}
=
F_j(R)\,
\overline G^{\,\mathrm{exc}}_j(R),
\label{eq:burden}
\end{equation}
for $N_j(R)>0$. The decomposition is useful because two trips can have the same burden for different reasons: one may contain many moderate anomalous episodes, while another contains fewer but statistically stronger departures. If no episode is detected, $F_j(R)$ and $B_j(R)$ are defined to be zero, while the episode-conditional quantity $\overline G^{\,\mathrm{exc}}_j(R)$ is left undefined.

The exposure adjustment prevents a longer trip from appearing more abnormal simply because it contains more observations. We write the reference explicitly because changing $R$ can change both the selected episodes and their statistical evidence. $B_j(R)$ therefore measures accumulated episode-level model mismatch per hour relative to a stated reference. Because $G(e;R)$ and $\lambda_j(R)$ are dimensionless divergence quantities, the only unit in $B_j(R)$ comes from the exposure denominator; we therefore interpret it as \emph{excess divergence per driving hour}, rather than as a physical unit such as acceleration or distance.

\paragraph{Occupied anomalous time.}
The non-overlap rule also gives a direct measure of how much of a trip is covered by detected anomalous episodes. If $d_e$ is the duration of episode $e$ in seconds, define
\begin{equation}
O_j(R)
=
\frac{1}{3600E_j}
\sum_{e\in\widehat{\mathcal E}_j(R)} d_e.
\label{eq:occupied-fraction}
\end{equation}
Because the selected episode cores do not overlap, $0\le O_j(R)\le1$, and $100O_j(R)\%$ is the percentage of valid driving time classified as anomalous by the interval procedure. Unlike episode frequency, occupied time reflects both the number and the duration of the selected episodes. We set $O_j(R)=0$ when no episode is detected.

\subsection{Signal-based characteristics of individual episodes}
\label{subsec:physical}

The statistical summaries above quantify departure from the fitted HMM but do not describe the corresponding vehicle movement. Once an episode has been localized, we return to the original lateral and longitudinal acceleration observations within that interval. We retain its duration $d_e$ and summarize both the magnitude and temporal allocation of the acceleration activity.

\paragraph{Acceleration magnitude.}
To describe how large the movement is within an episode, we use its root mean square (RMS) and peak bivariate acceleration magnitude,
\begin{equation}
A_e^{\mathrm{RMS}}
=
\left\{
\frac{1}{n_e}
\sum_{t\in e}\|Y_{jt}\|^2
\right\}^{1/2},
\qquad
A_e^{\mathrm{peak}}
=
\max_{t\in e}\|Y_{jt}\|,
\label{eq:amplitude}
\end{equation}
where $\|\cdot\|$ denotes the Euclidean norm, so
$\|Y_{jt}\|=\sqrt{(a^\lat_{jt})^2+(a^\lon_{jt})^2}$.
$A_e^{\mathrm{RMS}}$ summarizes the typical acceleration magnitude over the episode, while $A_e^{\mathrm{peak}}$ records the largest observed magnitude.

\paragraph{Temporal concentration (TC).}
Magnitude alone does not describe how an episode unfolds. Two episodes can have similar overall acceleration but differ substantially in whether the activity is sustained throughout the interval or concentrated in relatively few observations. Let
\[
r_{et}=\|Y_{jt}\|,
\qquad
w_{et}=\frac{r_{et}}{\sum_{s\in e}r_{es}}.
\]
We define the temporal concentration of episode $e$ by
\begin{equation}
\mathrm{TC}_e
=
\frac{n_e\sum_{t\in e}w_{et}^{\,2}-1}{n_e-1},
\qquad
0\le \mathrm{TC}_e\le1.
\label{eq:tc}
\end{equation}
TC is zero when acceleration magnitude is distributed evenly across the episode and increases as more of the activity is concentrated in fewer observations. Multiplying the complete episode by a common positive constant does not change $\mathrm{TC}_e$, so TC captures a different property from RMS or peak magnitude. TC measures the unevenness of acceleration magnitude across observations, but not their temporal order; the full trajectory is considered separately through DTW. TC is also sensitive to episode duration: for a comparable degree of unevenness in acceleration magnitude, shorter episodes generally have higher TC. However, duration alone does not determine TC, and we examine this relationship further in \cref{subsec:TC_results}.

\begin{figure}[htbp]
\centering
\includegraphics[width=\textwidth]{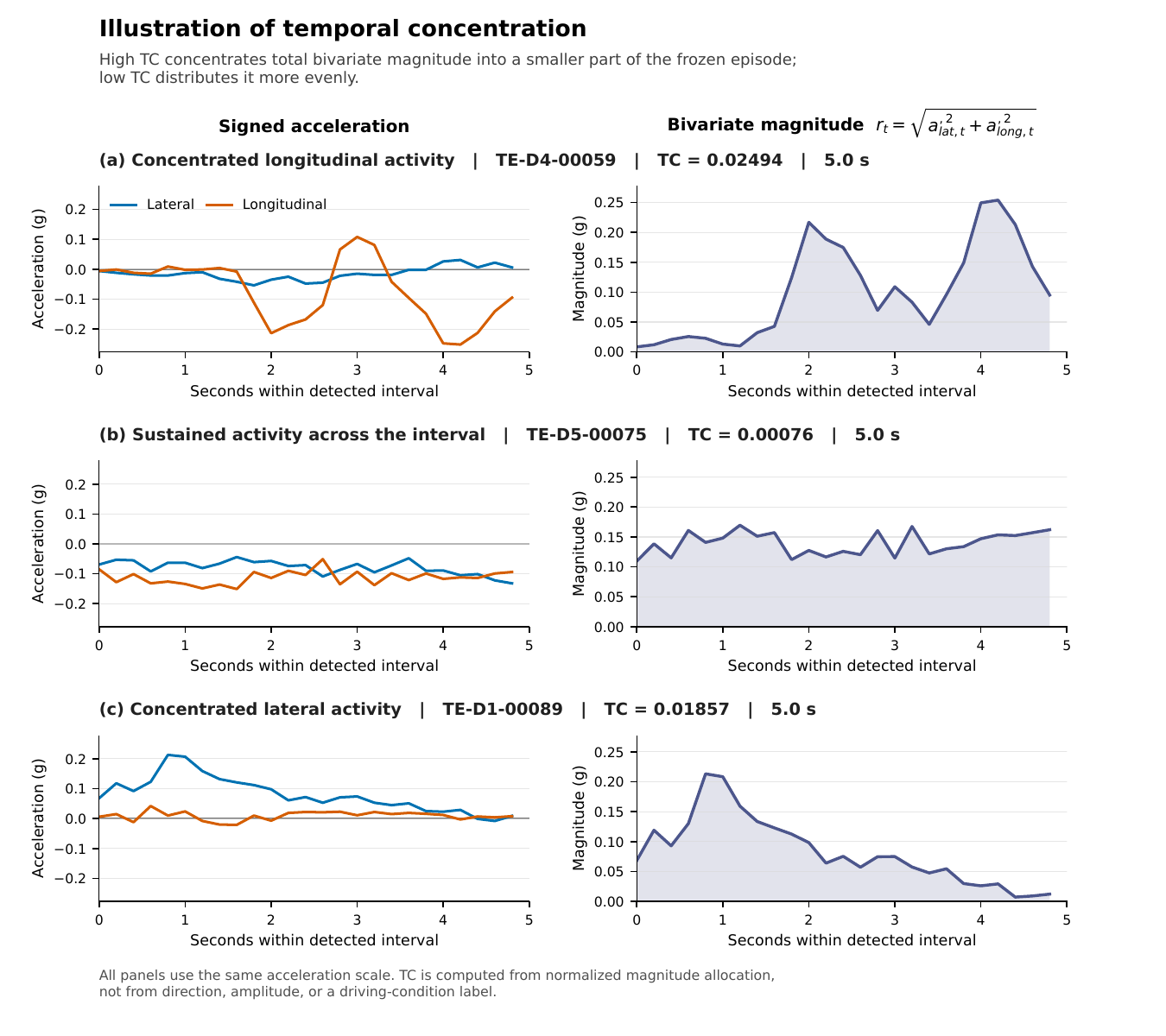}
\caption{Illustration of temporal concentration in the original acceleration signals using representative held-out episodes from the final pooled $K=15$, $q=1$ analysis. The concentrated longitudinal and lateral examples have TC values 0.02494 and 0.01857, respectively, compared with 0.00076 for the sustained example. All three episodes last five seconds and are displayed on the same acceleration scale. The examples were selected without using the driving-condition labels; TC describes the temporal allocation of acceleration magnitude rather than a predefined behaviour class.}
\label{fig:tc-physical-examples}
\end{figure}

\Cref{fig:tc-physical-examples} illustrates this distinction using three five-second held-out episodes from the empirical application described in \cref{sec:data}. In the concentrated examples, a small number of pronounced peaks or troughs in the signed acceleration signals produce corresponding peaks in the bivariate magnitude $r_{et}$. A relatively large share of the total activity is therefore concentrated around these times, giving a larger TC. In the sustained example, the bivariate magnitude remains comparatively even throughout the interval, with no similarly concentrated region, and TC is consequently close to zero.

For a trip with $N_j$ detected episodes, we summarize temporal concentration by
\begin{equation}
\overline{\mathrm{TC}}_j
=
\frac{1}{N_j}
\sum_{e\in\widehat{\mathcal E}_j}\mathrm{TC}_e.
\label{eq:trip-tc}
\end{equation}
Thus $\overline{\mathrm{TC}}_j$ is the mean TC across the detected episodes within trip $j$, not an average across different trips. It is defined only when at least one episode is detected, and the number of contributing episodes is retained when interpreting the trip-level mean.

\subsection{Recurring trajectory clusters and physical families}
\label{subsec:families}

RMS, peak magnitude, and TC summarize particular properties of an episode, but they do not describe its complete lateral--longitudinal trajectory. We therefore also group detected episodes with similar acceleration trajectories into recurring DTW clusters.

Each episode is represented by its complete bivariate acceleration trajectory. Before comparing shapes, both channels are divided by one common RMS amplitude. This allows episodes with a similar trajectory but different overall magnitude to be grouped together while preserving their relative lateral--longitudinal structure. The original magnitude remains available separately through \cref{eq:amplitude}.

\begin{figure}[htbp]
\centering
\includegraphics[width=\textwidth]{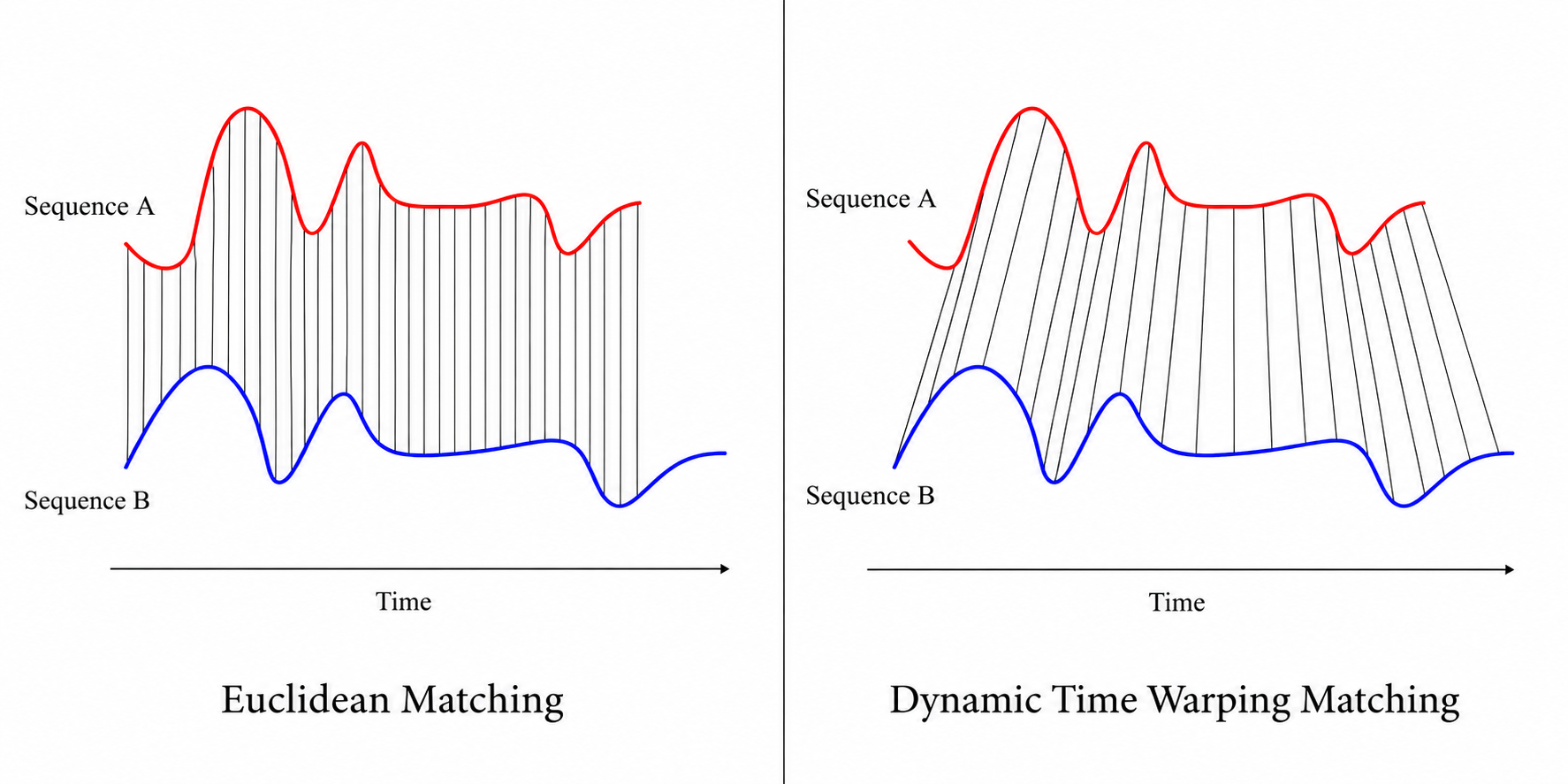}
\caption{Illustration of direct pointwise matching and dynamic time warping (DTW). Pointwise matching pairs observations according to their time positions, whereas DTW allows the alignment to stretch or compress locally while preserving temporal order.}
\label{fig:dtw-alignment}
\end{figure}

Similarity is measured using multivariate dynamic time warping (DTW), which allows episodes with different lengths or different local timing to be compared \citep{SakoeChiba1978}. \Cref{fig:dtw-alignment} illustrates the basic idea using two univariate sequences. Direct pointwise matching pairs observations according to their time positions and can therefore misalign otherwise similar features when one occurs earlier or later. DTW instead finds an order-preserving alignment that can stretch or compress locally, allowing corresponding peaks, troughs, and other features to be matched more closely. In our application, each observation is a bivariate lateral--longitudinal acceleration vector. Euclidean distance between these vectors is used as the local dissimilarity, and both acceleration channels share the same DTW alignment rather than being warped separately.

The resulting pairwise DTW distances are used to group recurring trajectories by complete-link clustering, with each cluster represented by an observed medoid. Once the cluster structure has been learned from a training sample, a new episode, such as one detected from a new trip, is compared with the learned cluster medoids. It is assigned to the nearest cluster when its DTW distance to that cluster's medoid is no greater than the fold-specific tolerance $\tau_r$ used for complete-linkage clustering (defined in \cref{sec:data}); otherwise, it is left unmatched.

\paragraph{Physical families.}
The recurring DTW clusters are data-dependent groups of similar trajectories rather than universal driving-behaviour classes. To make them easier to interpret and summarize, we assign each cluster to one of six detailed physical families using the original lateral and longitudinal acceleration trajectory of its medoid. The description reflects whether the movement is predominantly lateral or longitudinal, the direction of the longitudinal movement, and whether a lateral trajectory contains a substantial movement in the opposite direction. This produces the physical families \emph{longitudinal positive}, \emph{longitudinal negative/recovery}, \emph{longitudinal mixed}, \emph{lateral correction}, \emph{lateral excursion}, and \emph{mixed-axis}. For trip-level summaries, lateral correction and lateral excursion are combined into the broader \emph{lateral correction/excursion} family, leaving five broad physical families. Episodes that cannot be assigned to a recurring DTW cluster remain \emph{unmatched}.

The mapping uses only the acceleration trajectories and is fixed before the driving-condition labels are used for evaluation. It is intended as a simple description of recurring signal patterns, not as a claim that these physical families are universal or verified driving manoeuvres. The exact deterministic mapping is given in \cref{app:physical-mapping}.

\paragraph{Trip-level physical-family rates.}
For broad physical family $m$, we summarize recurrence in trip $j$ by
\begin{equation}
F_{jm}
=
\frac{N_{jm}}{E_j},
\label{eq:type-rate}
\end{equation}
where $N_{jm}$ is the number of episodes assigned to that family.

The resulting physical episode profile can be read in two complementary ways. First, \emph{occurrence} describes what kinds of episodes appear in a trip and how often they recur, through quantities such as $F_{jm}$ and the amount of exposure occupied by family $m$. Second, \emph{realization} describes how individual occurrences of that family are expressed through TC, RMS and peak magnitude, and duration. A trip can therefore differ from another because a particular pattern occurs more often, because similar episodes are realized more strongly or more unevenly, or both.

\section{Data and empirical design}
\label{sec:data}

\subsection{UAH-DriveSet and preprocessing}

We study the UAH-DriveSet, a publicly available controlled-driving dataset in which drivers were asked to mimic normal, aggressive, and drowsy driving on a motorway and a secondary road \citep{Romera2016}. We retain Drivers D1--D5, consistent with our earlier telematics studies \citep{Chan2025ASTIN}. Driver D6 is excluded because this driver used an electric vehicle and did not complete the full set of secondary-road trips owing to vehicle-autonomy constraints.

For each retained driver, the original data contain seven trips: one normal, one aggressive, and one drowsy motorway trip, together with two normal, one aggressive, and one drowsy secondary-road trips. The five-driver sample therefore contains 35 trips before data-quality screening.

The raw UAH-DriveSet records GPS-derived speed at 1 Hz and accelerometer signals at 10 Hz. We use the supplied Kalman-filtered lateral and longitudinal acceleration signals and place them on a regular 5 Hz grid by retaining the observation closest to each grid time. This preserves substantially more information about short acceleration patterns than the 1 Hz GPS series while reducing computation relative to the full 10 Hz recordings. Speed is not included in the present analysis because our focus is on short local acceleration episodes. Over intervals lasting only a few seconds, the 1 Hz speed series mainly provides broader driving context, whereas lateral and longitudinal acceleration more directly describe braking, acceleration, lateral correction, and recovery. This differs from \citet{Chan2025ASTIN}, where speed and the two acceleration channels were analysed jointly at 1 Hz.

\begin{figure}[p]
\centering
\includegraphics[width=0.92\textwidth]{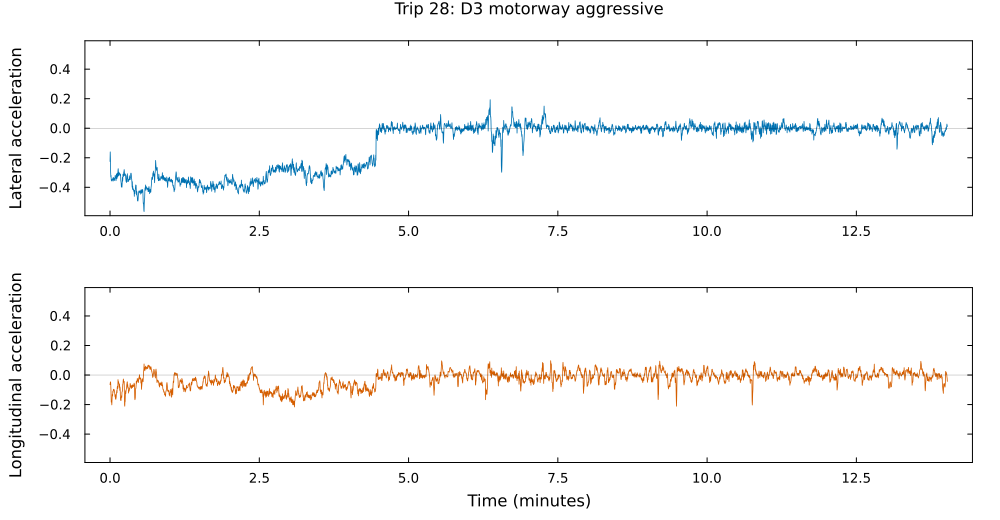}\\[0.8em]
\includegraphics[width=0.92\textwidth]{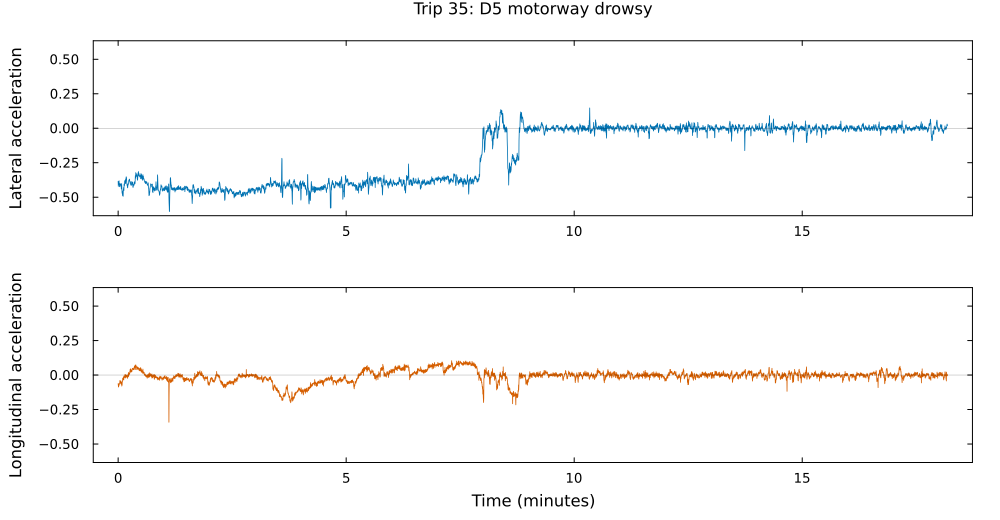}
\caption{Processed acceleration series for the two trips removed during data-quality screening. Trip 28 is the D3 motorway aggressive trip and Trip 35 is the D5 motorway drowsy trip. In both cases, lateral acceleration shows a sustained offset-like pattern followed by an abrupt return toward the usual range, while the longitudinal series also exhibits irregular behaviour around the affected portion.}
\label{fig:bad-trips}
\end{figure}

We inspect the processed acceleration signals before fitting the models. Trips 28 and 35 are excluded because their signal behaviour appears more consistent with a recording or preprocessing problem than with a plausible driving pattern. Trip 28 is the D3 motorway aggressive trip and Trip 35 is the D5 motorway drowsy trip. As shown in \cref{fig:bad-trips}, both contain a sustained offset-like pattern in lateral acceleration followed by an abrupt return toward the usual range. The longitudinal signal is also irregular around the affected portions, particularly for Trip 35. The issue therefore cannot be resolved reliably by simply recentering the lateral acceleration, since doing so would leave the joint bivariate trajectory distorted.

After these exclusions, the final analysis contains 33 trips: 15 normal, nine aggressive, and nine drowsy. There are 13 motorway trips and 20 secondary-road trips. In particular, D3 has no motorway aggressive trip and D5 has no motorway drowsy trip; no replacement trips or imputed comparisons are introduced.

\subsection{Evaluation and sensitivity design}
\label{subsec:evaluation-design}

\paragraph{Leave-one-driver-out construction.}
We use a leave-one-driver-out (LODO) design to examine whether the proposed episode-based trip measures, including statistical and physical summaries of the detected anomalous episodes in \cref{sec:measures}, transfer to a driver who was not used to construct them. In each fold, one driver is held out completely. The HMM, simulated residual reference, whole-search calibration, recurring DTW clusters, and their physical-family assignments are constructed using only the other four drivers and are then fixed before the held-out trips are evaluated. There is no driver-specific recentering, reference distribution, or scoring rule for the held-out driver.

\paragraph{Reference populations.}
Because the HMM-based anomaly measure is defined relative to a fitted reference model, we examine how the results change when the composition of that reference is varied. We consider the five reference populations in \cref{tab:reference-design}.

\begin{table}[htbp]
\centering
\caption{Reference populations used in the sensitivity analysis. The first four references contain 12 trips in every LODO fold and are balanced across the four training drivers.}
\label{tab:reference-design}
\begin{tabular}{ll}
\toprule
Reference & Training composition \\
\midrule
$R_N$ & 12 normal trips \\
$R_{N+A}$ & 8 normal and 4 aggressive trips \\
$R_{N+D}$ & 8 normal and 4 drowsy trips \\
$R_{N+A+D}$ & 4 normal, 4 aggressive, and 4 drowsy trips \\
$R_{\mathrm{pool}}$ & all eligible trips from the four training drivers \\
\bottomrule
\end{tabular}
\end{table}

For the first four references, each training driver contributes three trips, always including the normal motorway trip. When only one secondary-road normal trip is needed, we select the one with the lower trip ID. All aggressive and drowsy training trips are selected from the secondary road. The normal-only reference $R_N$ measures departure from dynamics learned from normal driving. The mixed references include abnormal driving conditions while retaining the same number of training trips, while $R_{\mathrm{pool}}$ uses all eligible trips from the four training drivers and contains 26 or 27 trips depending on the held-out driver. All reference populations pool motorway and secondary-road trips, reflecting our aim of constructing a common scoring framework across road types; road type is also controlled in the subsequent matched comparisons. The interval search and trip-level statistical summaries are otherwise unchanged, so these comparisons examine how the definition of ``unusual'' driving depends on the chosen reference population. The detailed DTW cluster and physical-family analysis below focuses on $R_{\mathrm{pool}}$, while all five reference populations are retained for the trip-level statistical and signal-based comparisons.

\paragraph{Analysis specification.}
The main specification uses a 15-state HMM and $q=1$, so each time point in an interval is represented by the current bivariate forecast residual only. We retain $K=15$ as the primary state resolution rather than interpreting it as a generally optimal number of HMM states. Candidate intervals contain 10, 15, \ldots, 50 observations, corresponding to durations of 2--10 seconds at 5 Hz, to capture short and longer anomalous episodes without imposing a single episode length.

For each fitted HMM, 500 complete trips are simulated to estimate the model-implied residual-reference moments $\widehat\mu_0$ and $\widehat\Sigma_0$ in \cref{eq:whiten}. The simulations do not all have the same length: an eligible observed trip design is sampled for each simulated trip, so trip length and valid residual structure vary across the simulation bank. The residual vectors from all 500 simulated trips are then pooled to estimate one common reference mean and covariance for that fitted HMM.

Whole-search calibration uses a separate simulation bank. For each required trip design, 1,999 independent simulated trips reproduce its usable residual horizon and candidate-interval search and are used to calibrate the trip-design-specific penalty $\lambda_j$ at the 5\% whole-search level, as described in \cref{subsec:intervals}. Unlike the common residual-reference moments, this calibration depends on the amount of searching permitted by the trip design.

For the DTW clustering analysis, the DTW alignment is restricted to a normalized-time band of 0.20, permitting moderate differences in local timing while avoiding unrestricted alignment. The resulting pairwise DTW distances are grouped using complete-linkage clustering. For two current clusters $A$ and $B$, their complete-linkage distance is
\[
D_{\mathrm{CL}}(A,B)
=
\max_{e\in A,\;f\in B}
d_{\mathrm{DTW}}(e,f).
\]
For each LODO fold $r$, we define a fold-specific DTW-distance tolerance $\tau_r$ as the 10th percentile of the eligible pairwise distances between training episodes from different trips. Two clusters can therefore be joined only when
\[
D_{\mathrm{CL}}(A,B)\leq\tau_r.
\]

Finally, a recurring DTW cluster must contain at least three episodes from at least two trips and two training drivers. This requirement is intended to retain patterns that recur beyond a single journey or individual rather than constructing clusters from repeated behaviour specific to one trip or driver.

\paragraph{Use of behaviour labels.}
Apart from defining the reference populations above, behaviour labels do not enter the detection, episode summaries, clustering, or held-out assignment procedures. They are used only after the trip-level outputs have been obtained to evaluate whether those outputs distinguish the instructed driving conditions and to fit the exploratory logistic regressions in \cref{subsec:auc-classification}. The fitted models can then be applied to new, unlabelled trips.

The trip labels describe the instructed condition of the complete trip and should not be interpreted as pointwise ground truth. A normal trip can contain an unusual manoeuvre, while an aggressive or drowsy trip can also contain periods of ordinary driving. The evaluation therefore concerns separation at the trip level rather than classification of every observation or detected episode.

\paragraph{Matched evaluation.}
The evaluation compares aggressive and drowsy trips with normal driving matched by both driver and road type. This focuses the comparison on differences associated with the instructed driving condition rather than persistent differences between drivers or road types.

Each driver has one normal motorway trip but two normal secondary-road trips, giving one matched comparator for each motorway aggressive or drowsy trip and two for each secondary-road aggressive or drowsy trip. After the exclusion of Trips 28 and 35, this gives 14 aggressive--normal and 14 drowsy--normal individual comparisons. For each trip-level measure defined in \cref{sec:measures}, we record how often the aggressive or drowsy trip has a higher value than its matched normal comparator.

For the secondary-road comparisons, we additionally require an aggressive or drowsy trip to exceed \emph{both} matched normal trips; motorway comparisons are unchanged because only one normal comparator is available. This gives a total of nine driver--road comparisons for aggressive driving and nine for drowsy driving.

Together, the individual and strict comparisons assess whether the proposed trip-level measures consistently distinguish the instructed abnormal conditions from comparable normal driving.

\section{Empirical results}
\label{sec:results}

We first validate the interval detector in controlled simulations where the location and form of the departure are known. We then apply the complete framework to the UAH-DriveSet, examining the statistical and physical trip-level measures and their sensitivity to the reference population. We next compare the trip-level discrimination of the episode measures with the original pointwise anomaly indices, before examining recurring physical patterns in the detected episodes.

\subsection{Detector validation}
\label{subsec:detector_validation}

Before interpreting the UAH-DriveSet episodes, we first examine whether the interval procedure can recover anomalous intervals when the underlying departure is known. We do not attempt to simulate realistic aggressive or drowsy driving. Instead, we introduce controlled changes directly on the standardized residual scale and compare the intervals subsequently selected by the detector with the known inserted intervals.

Sequences are generated under the fitted $K=15$ pooled HMM using a trip design selected uniformly from the 33 eligible observed trips and passed through the same forecast-residual construction used for the observed trips. One interval is then perturbed by changing its residual mean, variance, or contemporaneous covariance. Specifically, let $X_t=(X_{t,1},X_{t,2})^\top$ denote a standardized bivariate residual within the inserted interval. For a mean change, we add a shift of magnitude $\delta\in\{1,1.5\}$ to the first coordinate, the second coordinate, or both equally, using $(\delta,0)^\top$, $(0,\delta)^\top$, or $(\delta/\sqrt{2},\delta/\sqrt{2})^\top$, respectively. The factor $1/\sqrt{2}$ keeps the overall magnitude of the shift equal to $\delta$ in all three cases. For a variance change, we multiply both coordinates by $c\in\{0.5,1.5,2\}$, so their variances are scaled by $c^2$ relative to the reference. In particular, $c=0.5$ represents a decrease in variance to one-quarter of its reference value. For a covariance change, we replace $X_{t,2}$ with $\rho X_{t,1}+\sqrt{1-\rho^2}X_{t,2}$, where $\rho\in\{-0.7,0.7\}$, while leaving $X_{t,1}$ unchanged. This introduces a cross-channel correlation while preserving the marginal reference variances. All transformations are applied only within the inserted interval. The insertion location is chosen by first sampling an eligible residual segment uniformly and then a starting position within that segment, leaving at least three seconds of unperturbed observations on either side.

The detector is not told whether a departure has been inserted, where it occurs, or how many episodes it should return, and therefore still applies the complete interval search. For the results reported here, we focus on these 11 stronger perturbation settings across seven durations of 2, 3, 4, 5, 6, 8, and 10 seconds, giving 77 configurations. Each configuration is evaluated over 200 simulations, reusing the same baseline sequence for each replicate across configurations but selecting a new insertion location for each setting. We also use 500 simulations with no inserted departure to check the 5\% whole-search false-report level used to calibrate $\lambda_j$.

For simulations with an inserted interval, a departure is counted as detected if at least one selected episode overlaps it. We measure localization using intersection-over-union (IoU),
\[
\operatorname{IoU}(I,\widehat I)
=
\frac{|I\cap\widehat I|}
{|I\cup\widehat I|},
\]
where $I$ is the inserted interval and $\widehat I$ is a detected interval. An IoU of one means that the interval is recovered exactly, while values closer to zero indicate less overlap. If several episodes are selected, we use the largest IoU with the inserted interval. Reported median IoUs are calculated among simulations in which the inserted interval is detected.

\Cref{tab:detector_validation} summarizes the main pattern. Under the null, at least one episode is reported in 4.6\% of the 500 simulations, close to the intended 5\% whole-search false-report level. For inserted departures, detection increases substantially with duration. This is particularly clear for mean and variance changes, while covariance-only changes remain more difficult. Importantly, when an interval is detected, its location is generally recovered closely: median IoU is high for the stronger mean departures and improves with duration for the variance and covariance changes. No detected strong departure was split into multiple selected episodes at any tested duration.

\begin{table}[htbp]
\centering
\caption{Detection and localization of strong inserted departures under the $K=15$, $q=1$ detector at three representative durations. Detection requires at least one selected episode to overlap the inserted interval. Median best IoU is calculated among detected cases. The complete validation also considers durations of 3, 4, 6, and 8 seconds.}
\label{tab:detector_validation}
\begin{tabular}{lrrrrrr}
\toprule
& \multicolumn{2}{c}{2 seconds} & \multicolumn{2}{c}{5 seconds} & \multicolumn{2}{c}{10 seconds} \\
\cmidrule(lr){2-3}\cmidrule(lr){4-5}\cmidrule(lr){6-7}
Departure type & Detection & IoU & Detection & IoU & Detection & IoU \\
\midrule
Mean shift & 20.3\% & 1.000 & 78.1\% & 0.923 & 98.8\% & 0.961 \\
Variance change & 17.5\% & 0.667 & 61.3\% & 0.833 & 94.7\% & 0.961 \\
Covariance-only change & 0.0\% & --- & 10.0\% & 0.744 & 72.5\% & 0.923 \\
All strong departures & 15.9\% & 0.818 & 61.1\% & 0.852 & 92.9\% & 0.961 \\
\bottomrule
\end{tabular}
\end{table}

The duration pattern is consistent with the intended role of the interval score. The procedure is designed to accumulate evidence across residuals that may not be individually extreme but, when considered together, show an unusual level, variability, or cross-channel relationship. Short departures provide less opportunity for such evidence to accumulate, while covariance estimation is also less stable over short intervals and is more strongly shrunk toward the reference covariance $I_2$. Thus the interval detector is most effective for sustained collective departures rather than isolated extreme observations.

This distinction is supported by comparison with the pointwise anomaly detection in \citet{Chan2025ASTIN}, where an observation is identified as anomalous when the absolute forecast pseudo-residual exceeds three in at least one dimension. For this comparison, the pointwise detector identifies an inserted departure if at least one observation within that interval exceeds the $\pm3$ threshold in either channel, while the interval detector identifies it if at least one selected episode overlaps the inserted interval. Applying the same $|z|\ge3$ rule to the simulated departures shows that the pointwise detector contributes more unique detections at short durations, whereas the interval detector increasingly contributes unique detections as the departure persists.

Among strong two-second departures, 12.5\% are detected by both methods, 22.8\% only by the pointwise detector, 3.4\% only by the interval detector, and 61.3\% by neither. For ten-second departures, the corresponding proportions are 65.7\%, 2.6\%, 27.2\%, and 4.5\%. At each duration, these percentages are calculated over 2,200 inserted-departure evaluations (11 perturbation settings with 200 replicates each). The high proportion missed by both methods at two seconds reflects the difficulty of detecting short changes, which may neither produce individually extreme pseudo-residuals nor accumulate sufficient evidence to be detected collectively. The two procedures are not calibrated to a common whole-search error rate, so this comparison is intended to demonstrate complementary coverage rather than relative error-controlled power.

We further examine whether two nearby departures can be distinguished. Two strong four-second departures with joint mean shifts of magnitude $\delta=1.5$ are inserted with gaps of 0, 1, 2, 3, 5, or 8 seconds, with 200 simulations for each gap. We count a result as merged if one selected episode overlaps both inserted departures, and as correctly separated if the two departures are identified by distinct selected episodes, with neither selected episode overlapping the other departure. When the departures are adjacent or separated by only 1--2 seconds, 93.7\% are represented as one episode. A three-second gap is transitional, with 63\% correctly separated and 34\% merged. At gaps of 5 or 8 seconds, 91\% are correctly separated.

The transition around a three-second gap may be partly explained by the candidate interval lengths considered by the detector. Although both inserted departures last four seconds, the detector searches all candidate lengths from 2 to 10 seconds. Two four-second departures separated by a two-second gap span ten seconds and can therefore be contained within a single candidate interval. With a three-second gap, their combined span increases to eleven seconds, so no single candidate can fully contain both departures. However, a selected interval may still overlap both, and the final selection also depends on the penalized objective in \cref{eq:episode-selection}. The penalty discourages unnecessary splitting, while two sufficiently separated departures can be retained as distinct episodes when there is enough evidence.

Overall, the simulations verify the intended use of the proposed detector. It controls false reports under the fitted HMM, identifies and localizes sustained departures in the residual distribution, and converts nearby deviations into distinct episodes when there is sufficient evidence to separate them. Its weaker sensitivity to very short departures is a limitation, but also reflects its complementary purpose: pointwise anomaly detection identifies individually extreme residuals, while the interval detector is designed to identify residual patterns that become unusual collectively.

\subsection{Statistical anomaly burden and reference dependence}
\label{sec:reference_results}

Having validated the interval detector, we now examine the detected episodes in the observed UAH-DriveSet trips. We begin with their statistical evidence on the pseudo-residual scale. The exposure-adjusted statistical anomaly burden $B_j(R)$ in \cref{eq:burden} measures the accumulated excess interval evidence per driving hour. Because both the selected episodes and their excess evidence are defined relative to the fitted HMM, we examine how this quantity changes across the five reference populations in \cref{tab:reference-design}.

\Cref{tab:burden_reference} reports the matched abnormal--normal comparisons. The pairwise columns count the 14 aggressive--normal or drowsy--normal pairs in which the abnormal-condition trip has the larger burden. Under the stricter driver--road criterion, the abnormal trip must exceed every matched normal trip from the same driver and road type, giving nine strata for each abnormal condition.

\begin{table}[htbp]
\centering
\caption{Matched comparisons for exposure-adjusted statistical anomaly burden $B_j(R)$ in \cref{eq:burden}. Margins are abnormal minus matched normal and are summarized over the 14 individual comparisons.}
\label{tab:burden_reference}
\begin{tabular}{lrrr rrr}
\toprule
& \multicolumn{3}{c}{Aggressive} & \multicolumn{3}{c}{Drowsy} \\
\cmidrule(lr){2-4}\cmidrule(lr){5-7}
Reference
& Pairwise /14 & Strict /9 & Median margin
& Pairwise /14 & Strict /9 & Median margin \\
\midrule
$R_N$ & 14 & 9 & 2210.8 & 12 & 7 & 627.8 \\
$R_{N+A}$ & 12 & 7 & 857.1 & 7 & 4 & 143.3 \\
$R_{N+D}$ & 12 & 7 & 838.1 & 8 & 5 & 110.9 \\
$R_{N+A+D}$ & 12 & 7 & 683.6 & 8 & 4 & 121.4 \\
$R_{\mathrm{pool}}$ & 10 & 6 & 652.0 & 8 & 4 & 171.3 \\
\bottomrule
\end{tabular}
\end{table}

The normal-only reference gives the clearest separation. Aggressive trips have larger burden in 14/14 matched comparisons and 9/9 strict strata, while the corresponding drowsy results are 12/14 and 7/9. The ordering becomes less consistent under the broader references. Under $R_{\mathrm{pool}}$, the results fall to 10/14 and 6/9 for aggressive driving and 8/14 and 4/9 for drowsy driving. The median aggressive--normal margin also falls from 2210.8 under $R_N$ to 652.0 under $R_{\mathrm{pool}}$, while the drowsy--normal margin falls from 627.8 to 171.3. Statistical anomaly burden therefore contains useful abnormal--normal information, but the strength and consistency of that information depend substantially on the population used to define the HMM reference.

This reference effect is not simply a consequence of selecting many more or fewer episodes. Each reference is evaluated on the same 33 trips, with every trip scored in the leave-one-driver-out fold in which its driver is held out. \Cref{tab:reference_episodes} shows that the resulting episode selections are similar in overall size and occupied fraction, even though their accumulated statistical evidence differs considerably. For trip $j$, the occupied fraction in \cref{eq:occupied-fraction} is the proportion of total driving time covered by all selected episodes in that trip. The total excess evidence is the sum of $G(e;R)-\lambda_j(R)$ over all selected episodes across the 33 trips, or equivalently $\sum_{j=1}^{33}E_j B_j(R)$.

\begin{table}[htbp]
\centering
\caption{Selected episodes, occupied time, and accumulated excess evidence across reference populations for the same 33 trips. Mean occupied fraction is the average proportion of valid trip time covered by detected episodes.}
\label{tab:reference_episodes}
\begin{tabular}{lrrr}
\toprule
Reference & Detected episodes & Mean occupied fraction & Total excess evidence \\
\midrule
$R_N$ & 529 & 0.125 & 17135.5 \\
$R_{N+A}$ & 484 & 0.131 & 11812.5 \\
$R_{N+D}$ & 532 & 0.128 & 9973.9 \\
$R_{N+A+D}$ & 466 & 0.131 & 9547.0 \\
$R_{\mathrm{pool}}$ & 462 & 0.127 & 7782.7 \\
\bottomrule
\end{tabular}
\end{table}

For example, $R_{N+D}$ selects 532 episodes, slightly more than the 529 selected under $R_N$, yet total excess evidence is much smaller. Across all five references, the selected episodes together occupy, on average, approximately 12.5--13.1\% of valid driving time. Changing the reference therefore affects primarily how strongly the selected intervals depart from the fitted dynamics, rather than simply how many intervals are selected or how much of the trip they occupy.

The mixed-reference results also do not suggest a simple behaviour-specific accommodation effect: replacing normal training trips with aggressive trips weakens both aggressive and drowsy contrasts, and replacing them with drowsy trips does the same. The effect of including an abnormal driving condition in the reference is therefore not confined to that condition. More generally, broadening the population used to fit the HMM changes the range of driving dynamics that the model can accommodate, so departures that are pronounced relative to normal-only driving can appear less unusual relative to a more heterogeneous reference. The first four references contain the same number of training trips, whereas $R_{\mathrm{pool}}$ contains 26 or 27 trips per fold. Therefore, comparisons involving the pooled reference also reflect differences in training sample size, which cannot be separately assessed with the present dataset.

The reference population is therefore part of the definition of the statistical anomaly measure. In practice, a pooled portfolio reference is a natural default and measures what driving is unusual relative to the insurer's current portfolio. Such deviation can be related to risk, but it is not itself a definition of risky driving. Evaluating which reference is most useful for insurance risk assessment requires claim-linked data, as discussed in \cref{sec:conclusion}.

Overall, statistical anomaly burden provides evidence that aggressive and drowsy trips can depart from comparable normal driving, but it is neither equally consistent for the two abnormal conditions nor stable to the choice of reference population. We therefore examine the original acceleration signals to determine what additional information the detected episodes contain and whether signal-based characteristics can distinguish abnormal from matched normal driving more consistently.

\subsection{Signal-based characteristics of detected episodes}
\label{subsec:TC_results}

We next return to the original acceleration signals within the detected episodes. We compare trip-level mean temporal concentration (TC) in \cref{eq:trip-tc} with the RMS and peak acceleration magnitudes in \cref{eq:amplitude}, and examine whether their abnormal--normal ordering is preserved across reference populations.

\Cref{tab:signal_reference} summarizes the matched results. Mean TC gives the clearest common ordering for aggressive and drowsy driving: it is higher in 12--13 of the 14 aggressive--normal comparisons and 12--14 of the 14 drowsy--normal comparisons across all five references. The numerical TC values remain reference-dependent because changing the HMM changes which episodes are selected; it is the abnormal--normal ordering that is comparatively stable.

\begin{table}[htbp]
\centering
\caption{Matched ordering of signal-based trip measures across reference populations. Each entry gives the number of positive individual abnormal--normal comparisons out of 14, with the number of strict driver--road strata out of nine in parentheses.}
\label{tab:signal_reference}
\begin{tabular}{lcc cc cc}
\toprule
& \multicolumn{2}{c}{Mean TC}
& \multicolumn{2}{c}{Mean RMS}
& \multicolumn{2}{c}{Mean peak} \\
\cmidrule(lr){2-3}\cmidrule(lr){4-5}\cmidrule(lr){6-7}
Reference
& Aggressive & Drowsy
& Aggressive & Drowsy
& Aggressive & Drowsy \\
\midrule
$R_N$
& 13/14 (8/9) & 14/14 (9/9)
& 13/14 (8/9) & 12/14 (7/9)
& 14/14 (9/9) & 13/14 (8/9) \\
$R_{N+A}$
& 12/14 (7/9) & 12/14 (7/9)
& 13/14 (8/9) & 11/14 (6/9)
& 13/14 (8/9) & 13/14 (8/9) \\
$R_{N+D}$
& 13/14 (8/9) & 14/14 (9/9)
& 14/14 (9/9) & 11/14 (7/9)
& 14/14 (9/9) & 13/14 (8/9) \\
$R_{N+A+D}$
& 12/14 (7/9) & 12/14 (7/9)
& 12/14 (7/9) & 10/14 (6/9)
& 13/14 (8/9) & 11/14 (7/9) \\
$R_{\mathrm{pool}}$
& 13/14 (8/9) & 14/14 (9/9)
& 14/14 (9/9) & 11/14 (7/9)
& 13/14 (8/9) & 12/14 (7/9) \\
\bottomrule
\end{tabular}
\end{table}

Acceleration magnitude provides additional information, particularly for aggressive driving. Across the five references, RMS magnitude is higher in 12--14 of the 14 aggressive--normal comparisons and peak magnitude in 13--14 of 14, whereas the corresponding drowsy results are less consistent. TC is also empirically distinct from these amplitude measures: under the pooled reference, episode-level TC has Spearman correlations of only $0.104$ with RMS magnitude and $0.220$ with peak magnitude, and its correlation with episode duration is $-0.202$. These modest correlations suggest that TC is not simply a measure of acceleration magnitude or episode duration, although duration also enters its definition.

To examine the role of episode duration, let $\bar r_e$ denote the sample mean of acceleration magnitude within episode $e$, and define
\begin{equation}
V_e=\frac{\operatorname{Var}(r_{et})}{\bar r_e^2},
\qquad
\mathrm{TC}_e=\frac{V_e}{n_e-1},
\label{eq:tc-duration-relation}
\end{equation}
where $\operatorname{Var}(r_{et})$ is the population variance calculated across observations within the episode. Hence, $V_e$ is the squared coefficient of variation of acceleration magnitude, measuring its variability relative to the episode mean. As detailed in \cref{app:tc-duration}, the second expression is equivalent to the original definition of TC in \cref{eq:tc}. Thus, for two episodes with the same $V_e$, a two-second episode ($n_e=10$ at 5 Hz) has $(50-1)/(10-1)=49/9\approx5.44$ times the TC of a ten-second episode ($n_e=50$).

As a sensitivity check, we replaced TC with $V_e$ and averaged the resulting values across all detected episodes within each trip. Under $R_{\mathrm{pool}}$, aggressive--normal ordering falls from 13/14 to 3/14 matched comparisons, while drowsy--normal ordering falls from 14/14 to 11/14. Similar reductions in aggressive--normal separation occur under the other four references. These results indicate that the episode-length factor is important for TC's observed discriminatory performance, particularly for aggressive driving. Detailed results are reported in \cref{app:tc-duration}.

Taken together, mean TC provides relatively consistent separation between abnormal and matched normal driving across the five HMM references, while acceleration magnitude is more consistently elevated during aggressive driving than during drowsy driving.

\subsection{Classification using pointwise and episode-based measures}
\label{subsec:auc-classification}

We now evaluate whether the trip-level measures distinguish normal from abnormal driving more effectively than the original pointwise anomaly indices.

We consider an exploratory binary classification of the 33 retained trips, treating aggressive and drowsy driving as abnormal. For each of the five HMM reference populations, we compare three logistic regression models: the two lateral and longitudinal pointwise anomaly indices ($M_0$), trip-level mean TC alone ($M_{\mathrm{TC}}$), and their combination ($M_1$). Because of the small sample, we use Firth's bias-reduced logistic regression \citep{Firth1993}, which maximizes the penalized log-likelihood
\[
\ell_F(\boldsymbol{\beta})
=
\ell(\boldsymbol{\beta})
+
\frac{1}{2}\log\left|\mathcal I(\boldsymbol{\beta})\right|,
\]
where $\ell(\boldsymbol{\beta})$ is the usual logistic log-likelihood and $\mathcal I(\boldsymbol{\beta})$ is the Fisher information matrix. The penalty reduces small-sample bias and helps avoid unstable estimates under separation.

Following the LODO design, each regression is fitted using all eligible trips from the four training drivers and then used to predict the trips of the held-out driver. For a given reference and fold, the predictors for both training and held-out trips are calculated using the same fitted HMM and episode-detection procedure, while predictor standardization is based only on the regression training sample. The labels of normal, aggressive and drowsy are used to fit and evaluate the regressions, but do not enter the construction of the pointwise or episode-based measures, apart from defining the alternative HMM reference populations.

As in the preceding subsections, we first examine the matched driver--road comparisons. A comparison is successful when the abnormal trip receives a higher predicted probability than its matched normal trip. The stricter criterion requires the abnormal trip to exceed all matched normal trips from the same driver and road type. \Cref{tab:auc-matched} reports these comparisons for the three models. For $M_{\mathrm{TC}}$, the fitted TC coefficients are positive in all LODO folds, so the predicted probabilities preserve the matched ordering of mean TC reported in \cref{tab:signal_reference}.

\begin{table}[htbp]
\centering
\caption{Matched driver--road comparisons of held-out predicted probabilities under the three classification models. Entries report successful abnormal--normal comparisons out of 14, with successful strict driver--road strata out of nine in parentheses. The best results within each reference and driving condition are highlighted in bold.}
\label{tab:auc-matched}
\small
\begin{tabular}{lcccccc}
\toprule
& \multicolumn{3}{c}{Aggressive--normal} & \multicolumn{3}{c}{Drowsy--normal} \\
\cmidrule(lr){2-4}\cmidrule(lr){5-7}
Reference & $M_0$ & $M_{\mathrm{TC}}$ & $M_1$ & $M_0$ & $M_{\mathrm{TC}}$ & $M_1$ \\
\midrule
$R_N$ & \textbf{13/14 (8/9)} & \textbf{13/14 (8/9)} & 12/14 (7/9) & 11/14 (7/9) & \textbf{14/14 (9/9)} & \textbf{14/14 (9/9)} \\
$R_{N+A}$ & \textbf{12/14 (7/9)} & \textbf{12/14 (7/9)} & 11/14 (7/9) & 10/14 (5/9) & \textbf{12/14 (7/9)} & \textbf{12/14 (7/9)} \\
$R_{N+D}$ & 12/14 (8/9) & \textbf{13/14 (8/9)} & 12/14 (8/9) & 10/14 (6/9) & \textbf{14/14 (9/9)} & \textbf{14/14 (9/9)} \\
$R_{N+A+D}$ & \textbf{13/14 (8/9)} & 12/14 (7/9) & \textbf{13/14 (8/9)} & 10/14 (5/9) & \textbf{12/14 (7/9)} & 11/14 (6/9) \\
$R_{\mathrm{pool}}$ & \textbf{14/14 (9/9)} & 13/14 (8/9) & 12/14 (7/9) & 12/14 (7/9) & \textbf{14/14 (9/9)} & 13/14 (8/9) \\
\bottomrule
\end{tabular}
\end{table}

The TC-only model improves matched drowsy--normal ordering over the pointwise indices under every reference, reaching 14/14 comparisons under $R_N$, $R_{N+D}$, and $R_{\mathrm{pool}}$. For aggressive driving, the pointwise indices, particularly the longitudinal index, already give relatively strong separation, and TC alone does not consistently improve the matched ordering. Adding the pointwise indices to TC provides no systematic gain in these matched comparisons.

The matched comparisons focus on differences in driving condition for the same driver and road type. We next evaluate discrimination more broadly by calculating the area under the receiver operating characteristic curve (ROC-AUC) separately for each held-out driver and averaging the results across the five LODO folds. This includes comparisons across road types within each driver, while ensuring that all trips entering an AUC calculation are scored using the same fitted HMM and logistic regression. Each driver receives equal weight.

To evaluate the predicted probabilities themselves, we consider the Brier score, defined as $\mathrm{BS}=\frac{1}{33}\sum_{j=1}^{33}(\widehat p_j-y_j)^2$, where $\widehat p_j$ is the predicted probability of abnormal driving and $y_j$ is the corresponding binary label. The Brier score is a strictly proper scoring rule, with lower values indicating better probabilistic predictions. We also report log loss, which places a greater penalty on highly confident incorrect predictions. Unlike AUC, these two measures evaluate individual held-out probabilities and are calculated over all 33 trips.

\begin{table}[htbp]
\centering
\caption{Exploratory LODO classification of normal versus abnormal driving across the five HMM reference populations. AUC is averaged over the five held-out drivers, while Brier score and log loss use all 33 held-out trip probabilities. Higher AUC and lower Brier score and log loss indicate better performance. The best value for each measure within each reference is highlighted in bold.}
\label{tab:auc-classification}
\small
\begin{tabular}{llrrr}
\toprule
Reference & Model & Mean AUC & Brier score & Log loss \\
\midrule
$R_N$ & $M_0$ & 0.844 & 0.188 & 0.783 \\
& $M_{\mathrm{TC}}$ & \textbf{0.956} & \textbf{0.099} & \textbf{0.310} \\
& $M_1$ & 0.889 & 0.155 & 0.923 \\
\midrule
$R_{N+A}$ & $M_0$ & 0.783 & 0.195 & 0.892 \\
& $M_{\mathrm{TC}}$ & \textbf{0.878} & \textbf{0.126} & \textbf{0.503} \\
& $M_1$ & 0.833 & 0.186 & 0.782 \\
\midrule
$R_{N+D}$ & $M_0$ & 0.778 & 0.196 & 0.755 \\
& $M_{\mathrm{TC}}$ & \textbf{0.967} & \textbf{0.091} & \textbf{0.319} \\
& $M_1$ & 0.894 & 0.143 & 0.520 \\
\midrule
$R_{N+A+D}$ & $M_0$ & 0.806 & 0.179 & 0.633 \\
& $M_{\mathrm{TC}}$ & \textbf{0.844} & \textbf{0.134} & \textbf{0.404} \\
& $M_1$ & 0.806 & 0.158 & 0.681 \\
\midrule
$R_{\mathrm{pool}}$ & $M_0$ & 0.883 & 0.146 & 0.448 \\
& $M_{\mathrm{TC}}$ & \textbf{0.944} & 0.134 & 0.467 \\
& $M_1$ & 0.928 & \textbf{0.118} & \textbf{0.383} \\
\bottomrule
\end{tabular}
\end{table}

\Cref{tab:auc-classification} shows that among the three main models, TC alone gives the highest mean within-driver AUC and a lower Brier score than the original pointwise indices under all five HMM references. Under $R_{\mathrm{pool}}$, combining TC with the pointwise indices improves Brier score and log loss, but not mean AUC. While statistical anomaly burden $B_j(R)$ is sensitive to the reference population (\cref{sec:reference_results}), TC provides more consistent abnormal--normal ordering across references. Its classification advantage over the pointwise anomaly indices supports its use as a standalone behavioural measure for distinguishing the instructed driving conditions.

We also fitted two additional Firth models as sensitivity checks, using mean $V_e$ in \cref{eq:tc-duration-relation} and mean selected-episode duration instead of mean TC. Under $R_{\mathrm{pool}}$, replacing mean TC with mean $V_e$ reduces mean within-driver AUC from 0.944 to 0.522, indicating that the episode-length factor plays an important role in TC's discriminatory performance. However, mean TC also outperforms mean episode duration alone across all five HMM references, suggesting that it captures information beyond episode length. Detailed results are reported in \cref{tab:duration-classification} of \cref{app:tc-duration}.

These classification results remain exploratory because the study contains only 33 trips from five drivers. The hundreds of detected episodes are not independent driver-level observations, and the 14 matched comparisons for each abnormal condition reuse some normal trips. The reported AUCs and probability scores describe held-out predictions in this small controlled dataset, rather than established performance in a broader driving population or for insurance claims.

\subsection{Recurring episode clusters and physical families}
\label{subsec:family_results}

Having examined how well TC distinguishes normal from abnormal driving, we now go one step further to investigate the physical patterns of the detected episodes and understand how aggressive and drowsy driving differ. We group their lateral--longitudinal trajectories into recurring DTW clusters and assign these clusters to physical families.

Under the pooled $K=15$, $q=1$ specification, we construct recurring DTW clusters separately within each of the five LODO folds. In each fold, the clusters are constructed using detected episodes from the four training drivers. Episodes from the held-out driver are then assigned to the learned clusters based on their DTW distances to the cluster medoids, or left unmatched.

Across the five folds, we obtain 34--47 recurring DTW clusters per fold, giving 203 clusters in total. Of the 462 held-out episodes, 432 (93.5\%) are assigned to clusters learned from the other four drivers, while the remaining 30 are unmatched.

We then assign each DTW cluster to one of six detailed physical families using the original acceleration trajectory of its medoid, as described in \cref{subsec:families}. These families provide a coarser physical interpretation of the recurring clusters. For trip-level summaries, lateral correction and lateral excursion are combined into one family, leaving five broad physical families. \Cref{tab:physical_family_coarsening} shows how the 203 DTW clusters and their held-out assignments are distributed across the resulting five broad physical families.

\begin{table}[h!]
\centering
\caption{Distribution of fold-specific DTW clusters and their held-out assignments across the five broad physical families. The 30 unmatched episodes are excluded from the family counts.}
\label{tab:physical_family_coarsening}
\small
\begin{tabular}{lrr}
\toprule
Broad physical family & DTW clusters & Assigned held-out episodes \\
\midrule
Longitudinal positive & 17 & 60 \\
Longitudinal negative/recovery & 15 & 33 \\
Longitudinal mixed & 37 & 53 \\
Lateral correction/excursion & 63 & 193 \\
Mixed-axis & 71 & 93 \\
\bottomrule
\end{tabular}
\end{table}

\begin{figure}[h]
\centering
\includegraphics[width=\textwidth]{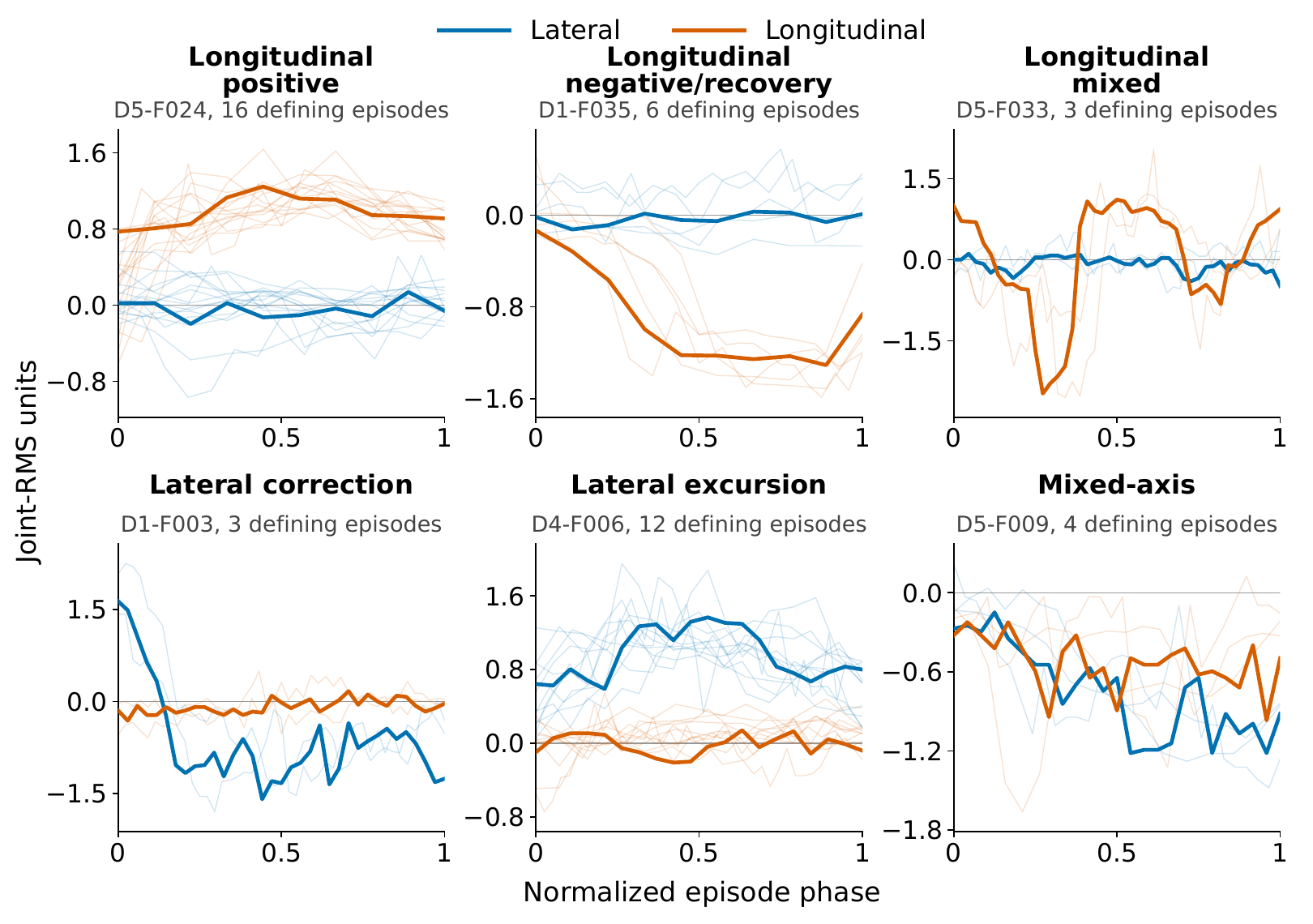}
\caption{Representative DTW clusters from the six detailed physical families under the pooled $K=15$, $q=1$ specification. Each panel shows the shape-normalized defining training episodes of one selected cluster in faint lines and its observed medoid in bold. Both acceleration channels are divided by the same episode-level RMS magnitude, and time is rescaled to normalized episode phase for display. Lateral correction and lateral excursion are subsequently combined for trip-level analysis.}
\label{fig:representative-families}
\end{figure}

\Cref{fig:representative-families} illustrates the six detailed physical families using one representative DTW cluster from each family. The longitudinal examples show a positive pattern, a negative excursion followed by recovery, and a mixed pattern with changes in longitudinal direction. The lateral-correction example contains a substantial movement in the opposite lateral direction, whereas the lateral-excursion example remains predominantly one-sided. The mixed-axis example contains substantial movement in both channels. These examples illustrate the physical descriptions but are not intended to represent every cluster within a family.

\begin{figure}[htbp]
\centering
\includegraphics[width=0.9\textwidth]{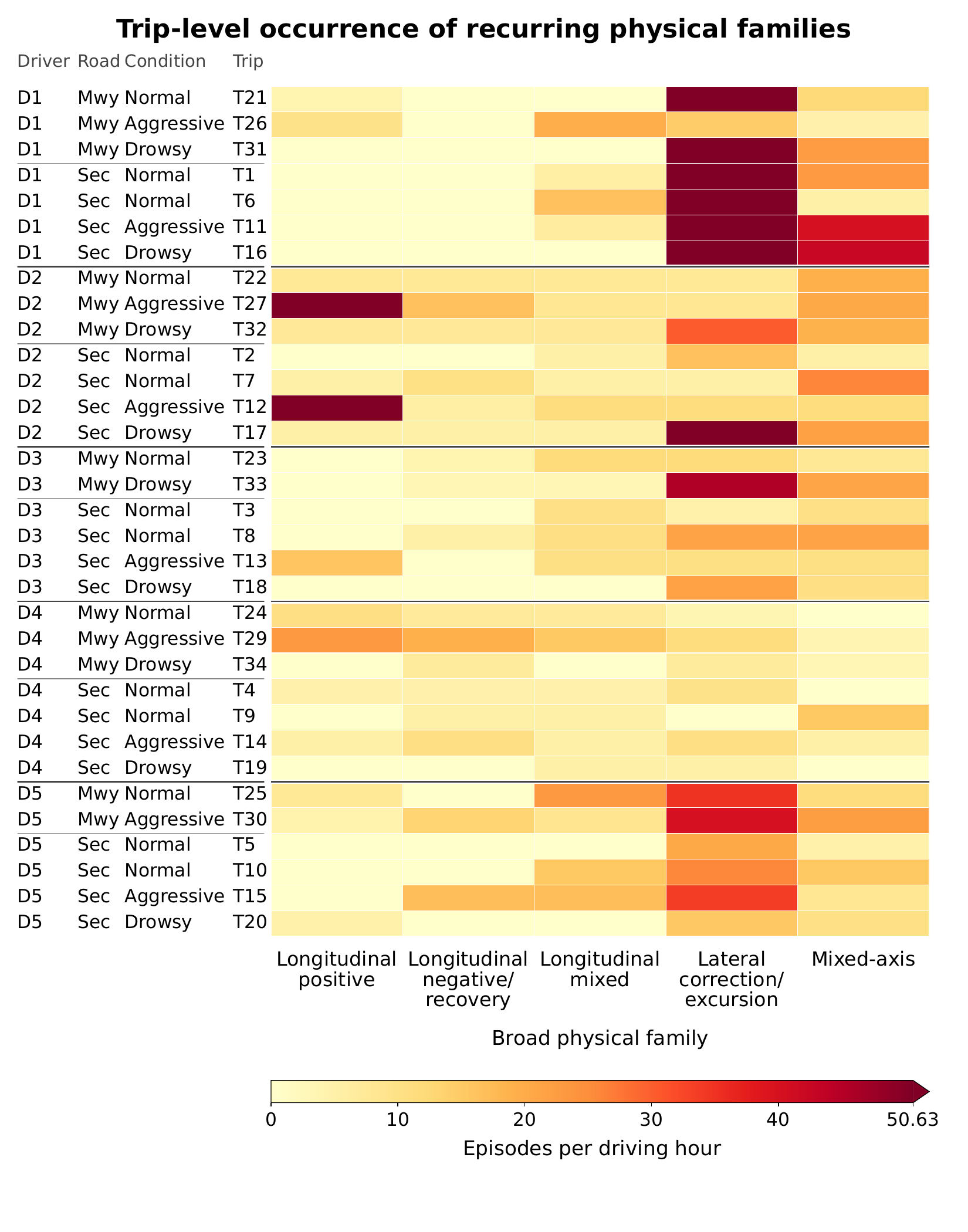}
\caption{Trip-level occurrence rates for the five broad physical families under the pooled reference. Rows preserve the matched driver, road, and driving-condition structure, while cells report episodes per driving hour. Colours are capped at the 95th percentile of the displayed rates (50.6 episodes per driving hour) to preserve visual resolution among moderate rates; nine cells exceed the cap and the underlying rates are unchanged. The DTW clustering and physical-family assignments do not use the driving-condition labels.}
\label{fig:trip_profile}
\end{figure}

\Cref{fig:trip_profile} shows the occurrence rates of the five broad physical families across all 33 trips. The substantial differences between drivers, including high lateral rates for several Driver D1 trips even under normal driving, reinforce the importance of comparing trips within the same driver and road setting rather than comparing absolute rates across drivers.

The trip-level profiles show different tendencies for aggressive and drowsy driving. For aggressive driving, longitudinal-positive episodes occur more often in 9/14 matched comparisons and 6/9 strict driver--road strata. When all three longitudinal families are combined, the mean peak and RMS magnitude of episodes belonging to these families are higher in the aggressive trip in all 13 comparable pairs. The remaining comparison, Aggressive Trip 15 versus Normal Trip 5 for Driver D5 on the secondary road, is undefined because Normal Trip 5 contains no detected longitudinal episode. Thus aggressive driving is associated not only with recurrence of longitudinal activity but also with greater acceleration magnitude when such episodes occur in both trips.

For drowsy driving, lateral correction/excursion episodes occur more often in 11/14 matched comparisons and 7/9 strict strata, and occupy more time in 9/14 comparisons. Thus the more consistent drowsy feature is repeated lateral activity rather than uniformly larger acceleration.

Finally, we use the matched motorway trips for Driver D2 to illustrate what the physical-family analysis reveals beyond trip-level mean TC. The mean TC across all detected episodes is 0.00730 for Normal Trip 22, compared with 0.01058 for Aggressive Trip 27 and 0.01178 for Drowsy Trip 32. Thus, mean TC already ranks both abnormal trips above the matched normal trip, but does not explain how their detected driving movements differ.

Aggressive Trip 27 contains 19 longitudinal-positive episodes, or 78.0 per driving hour, compared with two episodes, or 7.7 per hour, in Normal Trip 22. Their mean peak magnitude is also higher, at $0.105\,g$ versus $0.068\,g$. The additional information in this example therefore comes from much greater recurrence and amplitude of longitudinal-positive activity.

\begin{figure}[htbp]
\centering
\includegraphics[width=\textwidth]{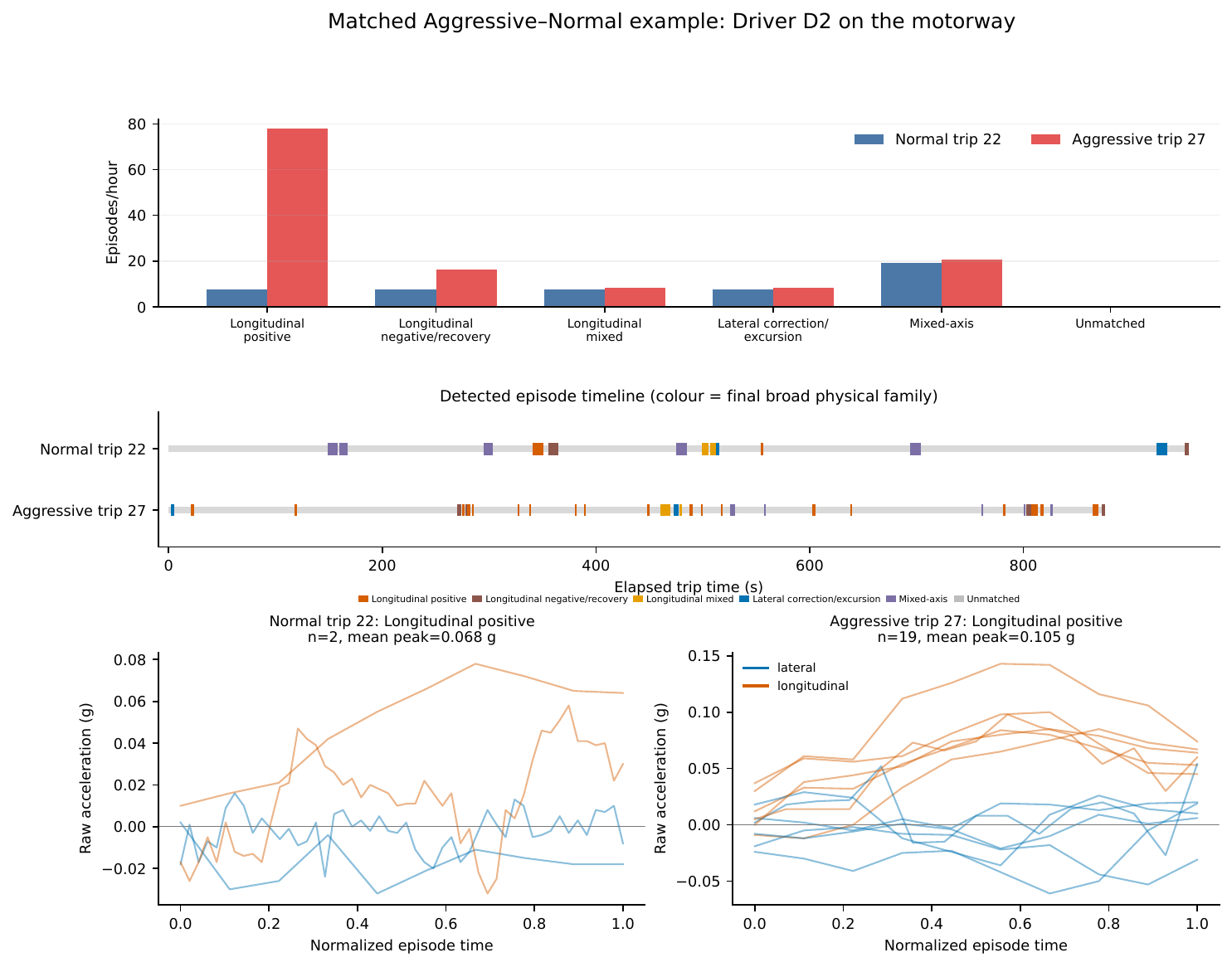}
\caption{Matched aggressive--normal comparison for Driver D2 on the motorway. Aggressive Trip 27 contains substantially more longitudinal-positive episodes than Normal Trip 22, and their mean peak acceleration is also higher.}
\label{fig:D2_aggressive}
\end{figure}

Drowsy Trip 32 instead contains eight lateral correction/excursion episodes, or 30.4 per driving hour, compared with two episodes, or 7.7 per hour, in Normal Trip 22. These episodes occupy substantially more driving time, although their mean peak magnitude is lower in this pair. Here the additional information comes primarily from the recurrence and occupancy of lateral activity.

\begin{figure}[htbp]
\centering
\includegraphics[width=\textwidth]{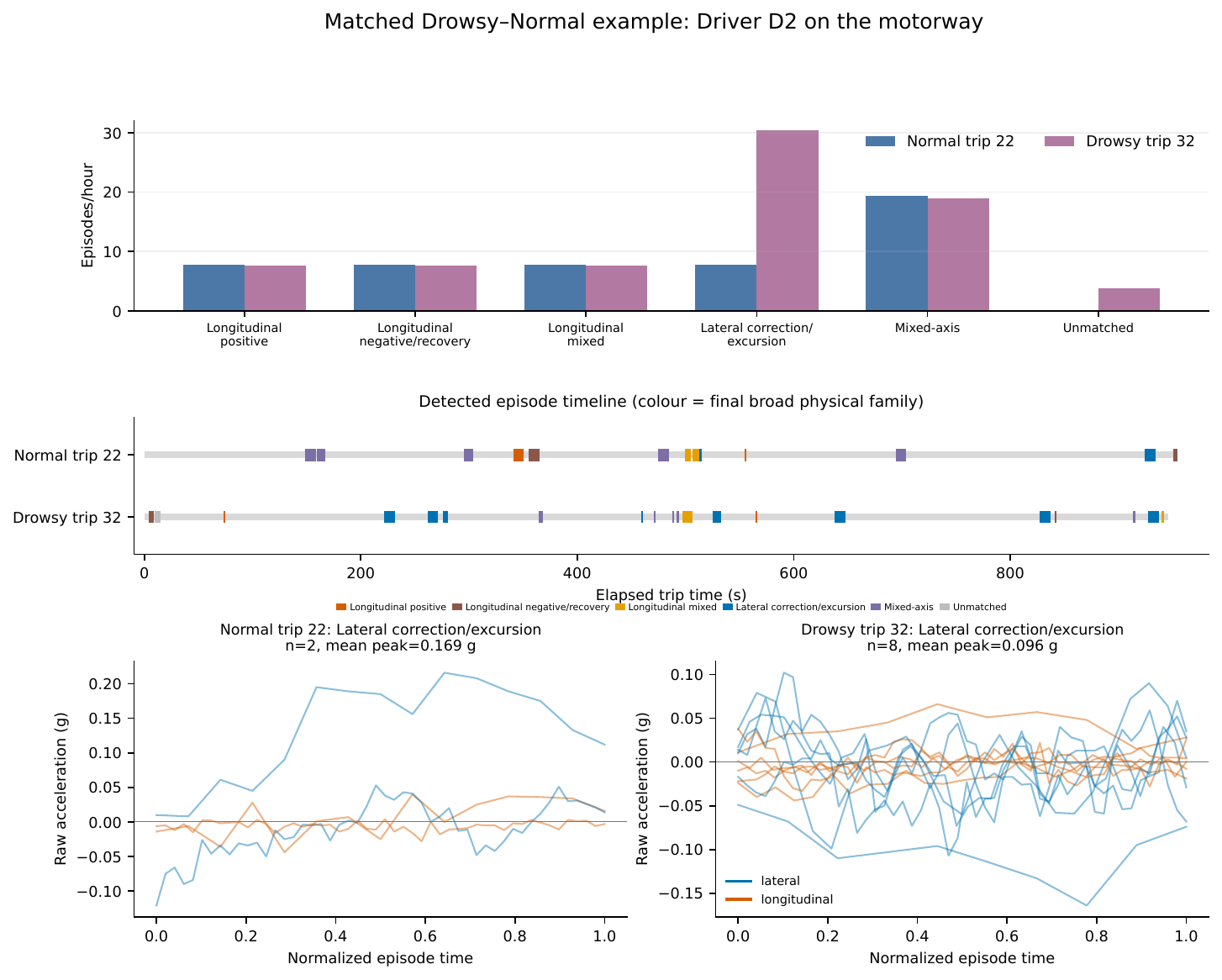}
\caption{Matched drowsy--normal comparison for Driver D2 on the motorway. Drowsy Trip 32 contains more lateral correction/excursion episodes than Normal Trip 22.}
\label{fig:D2_drowsy}
\end{figure}

Together, these two matched examples show what the physical-family profiles reveal beyond trip-level mean TC. While the aggressive trip shows both greater recurrence and magnitude of longitudinal episodes, the drowsy trip differs mainly through the greater recurrence of lateral activity. Thus, the physical-family profiles help distinguish how abnormal driving is expressed, even when both abnormal trips have higher overall TC than the matched normal trip. These physical-family measures could also be used in a second-stage classification to distinguish aggressive from drowsy driving. Given the limited data, we do not pursue this extension here.

\section{Conclusion and Actuarial Implications}
\label{sec:conclusion}

In this paper, we extend the HMM-based anomaly framework of \citet{Chan2025ASTIN} from pointwise anomalies to short anomalous episodes. Using forecast pseudo-residuals, we identify intervals of collective departure and return to the original acceleration signals to measure their temporal concentration and magnitude. Similar trajectories are further grouped into recurring DTW clusters and broader physical families, providing both statistical and physical descriptions of unusual driving behaviour.

We evaluate the framework using UAH-DriveSet, a controlled telematics dataset containing normal, aggressive, and drowsy driving. The results show that statistical anomaly burden depends substantially on the HMM reference population, whereas mean temporal concentration (TC) provides relatively consistent separation between abnormal and matched normal trips. In an exploratory leave-one-driver-out classification, mean TC alone achieves higher mean within-driver AUC and lower Brier score than the original pointwise anomaly indices across all five HMM references. The physical-family analysis complements these results by revealing different forms of abnormal driving: aggressive trips tend to exhibit stronger longitudinal acceleration, whereas drowsy trips show more recurring lateral corrections or excursions.

These findings are based on controlled driving conditions and do not yet establish a relationship with accident or insurance-claim risk. For actuarial applications, several questions remain concerning the choice of HMM reference population, the construction of driver-level behavioural summaries, and whether these measures improve the prediction of future insurance losses.

An important consideration is the choice of HMM reference population. A reference fitted to an insurer's current portfolio measures departures from the behaviour represented in that portfolio, while one constructed from drivers with favourable claim experience would provide a different benchmark. Neither is guaranteed to represent safe driving. Importantly, mean TC consistently distinguishes abnormal from normal driving across the five reference populations considered, suggesting that its discriminatory performance is robust. With larger claim-linked datasets, insurers could compare alternative references by examining the stability of the resulting behavioural measures and their predictive value for future losses.

For ratemaking, the trip-level measures would need to be summarized over a driver's repeated driving experience. As in \citet{Chan2025ASTIN}, upper quantiles of the distribution of a trip-level measure across a driver's repeated trips can capture occasional extreme behaviour that may be obscured by the average. Episode frequencies and physical-family occurrence rates can be expressed per unit of driving exposure, while TC and acceleration magnitudes describe the characteristics of the detected episodes. It is also important to retain the number of episodes contributing to these summaries, since a trip with no detected episode differs from one containing episodes with low TC. Using a common fitted HMM and physical-family dictionary would allow new trips to be evaluated on a consistent basis across drivers.

These driver-level summaries could then be evaluated alongside conventional rating variables, harsh-event measures and the original pointwise anomaly indices. For example, TC could be considered as a candidate covariate in claim-frequency modelling, with driving exposure appropriately accounted for. The physical-family profiles and other episode characteristics could further be examined for their relationships with claim frequency, severity or different types of accidents. Two drivers with similar overall TC may exhibit different physical patterns, and these differences may have different insurance implications. Such profiles could also help distinguish different forms of abnormal driving and potentially inform more specific driver feedback or training, although these applications remain to be evaluated. Rather than combining the measures using arbitrary weights, their contributions to risk assessment should be estimated from insurance outcomes and evaluated on independent policyholders.

As new driving experience becomes available, the behavioural assessment could also be updated over time. Credibility or Bayesian updating could combine existing risk information with the accumulating driving experience of an individual policyholder, taking into account the amount of available exposure. This could be particularly relevant for usage-based insurance, where risk classification and pricing may respond to changes in observed driving behaviour.

We are interested in investigating these questions further using real-world telematics data linked to insurance claims, ideally in collaboration with insurers or telematics providers. We hope that these research directions will also be of interest to practitioners exploring how richer descriptions of driving behaviour can contribute to insurance risk assessment and ratemaking.

\section*{Code availability}
The Julia implementation of the HMM-based anomalous-interval detection procedure is available in the public HMMToolkit repository at \url{https://github.com/ianwengchan/HMMToolkit}.

\bibliographystyle{apalike}
\bibliography{variance_episode_draft}

\newpage
\appendix

\section{Detailed mapping from DTW clusters to physical families}
\label{app:physical-mapping}

For reproducibility, this appendix records the deterministic rule used to map each recurring DTW cluster to a descriptive physical family. Let $f$ denote a DTW cluster and let $a^{\lat}_{ft}$ and $a^{\lon}_{ft}$ denote the lateral and longitudinal acceleration values of its medoid. We first calculate the integrated absolute lateral, negative-longitudinal, and positive-longitudinal activity
\begin{equation}
L_f
=
\Delta t\sum_t |a^{\lat}_{ft}|,
\qquad
B_f
=
\Delta t\sum_t \max(-a^{\lon}_{ft},0),
\qquad
U_f
=
\Delta t\sum_t \max(a^{\lon}_{ft},0),
\label{eq:family-physical-components}
\end{equation}
where $\Delta t$ is the sampling interval. The common factor $\Delta t$ cancels when the corresponding shares are formed,
\begin{equation}
\ell_f
=
\frac{L_f}{L_f+B_f+U_f},
\qquad
b_f
=
\frac{B_f}{L_f+B_f+U_f},
\qquad
u_f
=
\frac{U_f}{L_f+B_f+U_f},
\label{eq:family-physical-shares}
\end{equation}
so that $\ell_f+b_f+u_f=1$. Thus $\ell_f$ measures the lateral share of the medoid's total acceleration activity, while $b_f$ and $u_f$ measure its negative- and positive-longitudinal shares.

For lateral-dominant clusters, we additionally define the positive and negative lateral peak magnitudes
\begin{equation}
p_f^+
=
\max_t\max(a^{\lat}_{ft},0),
\qquad
p_f^-
=
\max_t\max(-a^{\lat}_{ft},0).
\label{eq:family-lateral-peaks}
\end{equation}

The physical family is assigned using the ordered rule in \cref{tab:physical-mapping-rule}.

\begin{table}[htbp]
\centering
\caption{Deterministic mapping from a recurring DTW cluster to its descriptive physical family. Conditions are evaluated from top to bottom.}
\label{tab:physical-mapping-rule}
\small
\renewcommand{\arraystretch}{1.35}
\begin{tabular}{p{0.60\textwidth}p{0.35\textwidth}}
\toprule
Condition & Physical family \\
\midrule

$\ell_f\ge0.70$ and
$\displaystyle
\frac{\min(p_f^+,p_f^-)}
{\max(p_f^+,p_f^-)}
\ge\frac13$
&
Lateral correction
\\[3ex]

$\ell_f\ge0.70$ and
$\displaystyle
\frac{\min(p_f^+,p_f^-)}
{\max(p_f^+,p_f^-)}
<\frac13$
&
Lateral excursion
\\[3ex]

$\ell_f\le0.30$ and $b_f\ge0.60$
&
Longitudinal negative/recovery
\\[0.4ex]

$\ell_f\le0.30$ and $u_f\ge0.60$
&
Longitudinal positive
\\[0.4ex]

$\ell_f\le0.40$
&
Longitudinal mixed
\\[0.4ex]

Otherwise
&
Mixed-axis
\\
\bottomrule
\end{tabular}
\end{table}

The distinction between lateral correction and lateral excursion is retained when describing individual recurring DTW clusters, but the two are combined into a single \emph{lateral correction/excursion} family for the main trip-level analysis. Both represent strongly lateral trajectories, while their finer distinction depends only on the presence of a sufficiently large opposite-direction lateral component. Without additional information such as steering angle, lane position, or road geometry, we do not interpret this distinction as identifying two verified manoeuvre classes. Combining them therefore keeps the trip-level physical profile deliberately coarse and avoids placing undue emphasis on this finer descriptive split.

\section{Sensitivity of temporal concentration to episode duration}
\label{app:tc-duration}

This appendix provides the derivation of the relationship between temporal concentration (TC) and episode duration, together with the sensitivity results discussed in \cref{subsec:TC_results,subsec:auc-classification}. Throughout the analysis, we retain the same fitted HMMs, detected episodes, and leave-one-driver-out (LODO) evaluation design. Only the measures used to summarize the detected episodes are changed.

\paragraph{Relationship between TC and episode duration.}
Recall from \cref{eq:tc} that TC is defined using the proportion of total acceleration magnitude at each observation within an episode. For episode $e$ containing $n_e$ observations, let $r_{et}$ denote the bivariate acceleration magnitude at observation $t$, and let
\[
\bar r_e=\frac{1}{n_e}\sum_{t\in e}r_{et},
\qquad
\operatorname{Var}(r_{et})=\frac{1}{n_e}\sum_{t\in e}(r_{et}-\bar r_e)^2
\]
denote its sample mean and variance, respectively. Here, the variance is calculated using $n_e$ rather than $n_e-1$ in the denominator.

We define the squared coefficient of variation as
\[
V_e=\frac{\operatorname{Var}(r_{et})}{\bar r_e^2}.
\]
This measures the variability of acceleration magnitude relative to its mean. In particular, multiplying all acceleration magnitudes within an episode by the same positive constant does not change $V_e$.

From \cref{eq:tc}, the normalized magnitude weights satisfy
\[
w_{et}=\frac{r_{et}}{\sum_{s\in e}r_{es}}
=\frac{r_{et}}{n_e\bar r_e}.
\]
Therefore,
\begin{align*}
n_e\sum_{t\in e}w_{et}^2-1
&=\frac{1}{n_e\bar r_e^2}\sum_{t\in e}r_{et}^2-1\\
&=\frac{\frac{1}{n_e}\sum_{t\in e}r_{et}^2-\bar r_e^2}{\bar r_e^2}\\
&=\frac{\operatorname{Var}(r_{et})}{\bar r_e^2}\\
&=V_e.
\end{align*}
Substituting this into the original definition of TC gives
\[
\mathrm{TC}_e
=\frac{n_e\sum_{t\in e}w_{et}^2-1}{n_e-1}
=\frac{V_e}{n_e-1},
\]
which establishes the relationship in \cref{eq:tc-duration-relation}.

Thus, TC measures relative variability in acceleration magnitude, with an additional factor depending on the number of observations in the episode. For example, at 5 Hz, a two-second episode contains 10 observations and a ten-second episode contains 50 observations. If the two episodes have the same $V_e$, their TC values differ by
\[
\frac{\mathrm{TC}_{2\mathrm{s}}}{\mathrm{TC}_{10\mathrm{s}}}
=\frac{V_e/(10-1)}{V_e/(50-1)}
=\frac{49}{9}\approx5.44.
\]
As a simple illustration, repeating the same ten-observation acceleration-magnitude pattern five times produces a 50-observation sequence with the same mean and variance, and hence the same $V_e$, but a lower TC.

\paragraph{Sensitivity analysis.}
To investigate the empirical importance of this length factor, we consider three measures for each trip $j$ containing $N_j>0$ detected episodes:
\[
\overline{\mathrm{TC}}_j
=\frac{1}{N_j}\sum_{e\in\widehat{\mathcal E}_j}\mathrm{TC}_e,
\qquad
\overline V_j
=\frac{1}{N_j}\sum_{e\in\widehat{\mathcal E}_j}V_e,
\qquad
\overline d_j
=\frac{1}{N_j}\sum_{e\in\widehat{\mathcal E}_j}d_e.
\]
Here, $\overline{\mathrm{TC}}_j$ is the original mean TC, $\overline V_j$ is the mean squared coefficient of variation, and $\overline d_j$ is the mean duration of the detected episodes in seconds. All three measures are averaged over the same selected episodes, including those not assigned to a recurring DTW cluster.

Comparing mean TC with mean $V$ allows us to examine the effect of removing the explicit episode-length factor, while mean duration provides a further comparison to determine whether TC's performance can be explained by episode length alone.

We first examine the matched driver--road comparisons, following the evaluation design in \cref{subsec:evaluation-design}. A comparison is successful when the abnormal trip has a higher value than its matched normal trip. There are 14 individual comparisons for each abnormal driving condition and nine stricter driver--road strata, where a secondary-road abnormal trip must exceed both matched normal trips.

\Cref{tab:duration-sensitivity} reports the results across the five HMM references. Replacing mean TC with mean $V$ substantially weakens aggressive--normal separation under every reference. Under $R_{\mathrm{pool}}$, successful aggressive comparisons fall from 13/14 to 3/14, while the corresponding drowsy comparisons fall from 14/14 to 11/14. The drowsy results are generally less affected.

\begin{table}[htbp]
\centering
\caption{Matched abnormal--normal ordering using trip-level mean TC and mean $V$ across the five HMM reference populations. Entries report successful individual comparisons, with successful strict driver--road strata in parentheses. Mean $V$ is calculated by replacing each episode's TC with $V_e=(n_e-1)\mathrm{TC}_e$ before averaging within the trip.}
\label{tab:duration-sensitivity}
\setlength{\tabcolsep}{6pt}
\begin{tabular}{@{}lcccc@{}}
\toprule
& \multicolumn{2}{c}{Aggressive--normal} & \multicolumn{2}{c}{Drowsy--normal} \\
\cmidrule(lr){2-3}\cmidrule(l){4-5}
Reference & Mean TC & Mean $V$ & Mean TC & Mean $V$ \\
\midrule
$R_N$ & 13/14 (8/9) & 6/14 (3/9) & 14/14 (9/9) & 13/14 (8/9) \\
$R_{N+A}$ & 12/14 (7/9) & 2/14 (1/9) & 12/14 (7/9) & 13/14 (8/9) \\
$R_{N+D}$ & 13/14 (8/9) & 6/14 (3/9) & 14/14 (9/9) & 10/14 (5/9) \\
$R_{N+A+D}$ & 12/14 (7/9) & 2/14 (1/9) & 12/14 (7/9) & 12/14 (7/9) \\
$R_{\mathrm{pool}}$ & 13/14 (8/9) & 3/14 (2/9) & 14/14 (9/9) & 11/14 (7/9) \\
\bottomrule
\end{tabular}
\end{table}

We next compare the three measures through separate one-variable Firth logistic regression models, following the same LODO procedure described in \cref{subsec:auc-classification}. For each reference population, the regression is fitted using trips from four drivers and evaluated on the held-out driver. The response is the binary indicator of normal versus abnormal driving, where aggressive and drowsy trips are combined into the abnormal category.

We report the mean within-driver AUC, calculated by averaging the five held-out driver AUCs, and the Brier score across all 33 held-out trips. Unlike the matched comparisons above, these results evaluate the fitted classification models rather than comparing the trip-level measures directly.

\Cref{tab:duration-classification} shows that mean TC has higher AUC and lower Brier score than mean episode duration under all five HMM references. Under $R_{\mathrm{pool}}$, mean TC achieves AUC 0.944 and Brier score 0.134, compared with 0.856 and 0.177 for mean duration. In contrast, replacing mean TC with mean $V$ reduces AUC to 0.522 and increases the Brier score to 0.258. Similar deterioration is observed across the other references.

\begin{table}[htbp]
\centering
\caption{Exploratory LODO classification using one trip-level measure at a time. AUC is averaged across the five held-out drivers, while the Brier score is calculated over all 33 held-out trips. Higher AUC and lower Brier score indicate better performance. The best values within each reference are highlighted in bold.}
\label{tab:duration-classification}
\setlength{\tabcolsep}{5pt}
\begin{tabular}{@{}lrrrrrr@{}}
\toprule
& \multicolumn{2}{c}{Mean duration}
& \multicolumn{2}{c}{Mean TC}
& \multicolumn{2}{c}{Mean $V$} \\
\cmidrule(lr){2-3}\cmidrule(lr){4-5}\cmidrule(l){6-7}
Reference & AUC & Brier & AUC & Brier & AUC & Brier \\
\midrule
$R_N$ & 0.811 & 0.192 & \textbf{0.956} & \textbf{0.099} & 0.639 & 0.227 \\
$R_{N+A}$ & 0.800 & 0.188 & \textbf{0.878} & \textbf{0.126} & 0.589 & 0.239 \\
$R_{N+D}$ & 0.822 & 0.204 & \textbf{0.967} & \textbf{0.091} & 0.383 & 0.264 \\
$R_{N+A+D}$ & 0.806 & 0.167 & \textbf{0.844} & \textbf{0.134} & 0.422 & 0.257 \\
$R_{\mathrm{pool}}$ & 0.856 & 0.177 & \textbf{0.944} & \textbf{0.134} & 0.522 & 0.258 \\
\bottomrule
\end{tabular}
\end{table}

Overall, these results show that TC's ability to distinguish abnormal from normal driving depends in part on its episode-length factor, particularly for aggressive driving. However, mean TC also performs better than mean episode duration alone, suggesting that its performance is not simply due to shorter detected episodes. Rather, TC reflects both the relative variability of acceleration magnitude and episode length. These findings are based on a small controlled dataset and warrant further investigation with more extensive driving and insurance data.

\end{document}